\documentclass[a4paper,fleqn]{cas-sc}

\usepackage{tablefootnote} 
\usepackage{threeparttable}

\usepackage{svg}
\usepackage[numbers]{natbib}
\usepackage[textsize=tiny]{todonotes}
\usepackage{soul}
\usepackage{multicol}
\usepackage{graphicx}

\usepackage{flushend}
\usepackage{soul}
\usepackage{hhline}
\usepackage{colortbl}
\usepackage{xcolor}
\usepackage{pdfpages}
\usepackage{textcomp}
\usepackage{multirow}
\usepackage{bigstrut}
\usepackage{array}
\usepackage{comment}
\usepackage{booktabs}
\renewcommand{\arraystretch}{1.5}
\usepackage[justification=centering]{caption}
\usepackage[justification=centering]{subcaption}
\usepackage{algorithm}
\usepackage{algpseudocode}
\usepackage[none]{hyphenat}
\def\tsc#1{\csdef{#1}{\textsc{\lowercase{#1}}\xspace}}
\tsc{WGM}
\tsc{QE}
\tsc{EP}
\tsc{PMS}
\tsc{BEC}
\tsc{DE}

\let\printorcid\relax

\begin{document}
\let\WriteBookmarks\relax
\def\floatpagepagefraction{1}
\def\textpagefraction{.001}

\shorttitle{\textit{ADAPTOR}}

\shortauthors{Ehsan Kabir et~al.}

\title [mode = title]{A Runtime-Adaptive Transformer Neural Network Accelerator on FPGAs}                      
\tnotemark[1]

\tnotetext[1]{This material is based upon work supported by the National Science Foundation under Grant No. 1956071.}


%
\author[1]{\color{black}Ehsan Kabir}

\cormark[1]

\fnmark[1]

\ead{ekabir@uark.edu}



\affiliation[1]{organization={Department of Electrical Engineering and Computer Science, University of Arkansas},
    city={Fayetteville},
    state={Arkansas},
    country={USA}}

\author[2]{\color{black}Jason D. Bakos}
\ead{jbakos@cse.sc.edu}
\affiliation[1]{organization={Department of Computer Science and Engineering, University of South Carolina},
    city={Columbia},
    state={South Carolina},
    country={USA}}
    
\author[1]{\color{black}David Andrews}
\ead{dandrews@uark.edu}

\author[1]{\color{black}Miaoqing Huang}
\ead{mqhuang@uark.edu}

\cortext[cor1]{Corresponding author}

\fntext[fn1]{This is the first author footnote.}


\begin{abstract}
Transformer neural networks (TNN) excel in natural language processing (NLP), machine translation, and computer vision (CV) without relying on recurrent or convolutional layers. However, they have high computational and memory demands, particularly on resource constrained devices like FPGAs. Moreover, transformer models vary in processing time across applications, requiring custom models with specific parameters. Designing custom accelerators for each model is complex and time-intensive. Some custom accelerators exist with no runtime adaptability, and they often rely on sparse matrices to reduce latency. However, hardware designs become more challenging due to the need for application-specific sparsity patterns. This paper introduces ADAPTOR, a runtime-adaptive accelerator for dense matrix computations in transformer encoders and decoders on FPGAs. ADAPTOR enhances the utilization of processing elements and on-chip memory, enhancing parallelism and reducing latency. It incorporates efficient matrix tiling to distribute resources across FPGA platforms and is fully quantized for computational efficiency and portability. Evaluations on Xilinx Alveo U55C data center cards and embedded platforms like VC707 and ZCU102 show that our design is 1.2$\times$ and 2.87$\times$ more power efficient than the NVIDIA K80 GPU and the i7-8700K CPU respectively. Additionally, it achieves a speedup of 1.7 to 2.25$\times$ compared to some state-of-the-art FPGA-based accelerators.
\end{abstract}




\maketitle

\section{Introduction}

Transformer neural networks (TNN) have shown great performance in natural language processing (NLP) \cite{language}, machine translation \cite{NMT}, computer vision \cite{wang_via_2022}, and other fields in recent years. While recurrent neural network (RNN) \cite{rnn_cho} and long short-term memory (LSTM) \cite{lstm_ho} models run sequential computation tasks during both training and inference, transformer facilitates high levels of computation parallelism throughout both processes using an attention mechanism. Thus, TNN is becoming a potential alternative to CNN, RNN, and LSTM \cite{self_img, self_cv}. 
There are many transformer models, such as full transformers containing both encoder and decoder \cite{attention}, BERT \cite{PretrainedTF, bert}, ALBERT \cite{albert}, structBERT \cite{structbert}, and others. 
These models contain different numbers of encoder and decoder stack \cite{attention} for different applications. A single encoder will often require a latency on the order of 100s of $\mu S$ \cite{peng_length_2022}. 
Around 38\% to 64\% of this time is spent in the multihead attention (MHA) mechanism depending on the number of tokens in the input sequence \cite{ham_elsa_2021, self_seq}, and the rest of the time is spent on feed forward network (FFN). Unfortunately, general-purpose platforms like GPUs and CPUs often suffer from low computational efficiency, underutilized memory bandwidth, and substantial compilation overheads for MHA layers \cite{flightLLM}. MHA and FFN \todo[disable]{FFN is defined and only use once, I would trim done the number of acronyms in the paper.} also occupy most of the on chip storage units \cite{mha_compute, li_ftrans_2020, qi_accommodating_2021}. \todo[disable]{what does this mean? are you talking about on-chip memory? \textcolor{red}{\textbf{\hl{(Ehsan's Ans.: Fixed)}}}}  
Therefore, it is essential to prioritize efficient hardware deployment on resource-constrained devices. 
FPGAs have gained widespread use for accelerating DNNs due to their high level of parallelism, high energy efficiency, and low latency \cite{dl, gan_hardware-aware_2022}. 
Recently, some works have successfully built FPGA based custom hardware accelerators for transformers \cite{lu_hardware_2020, peng_length_2022, li_ftrans_2020}. Application-specific integrated circuits (ASIC)-based accelerators also exist \cite{A3AA}. 

Lu et al. \cite{lu_hardware_2020} accelerated the attention mechanism and feedforward network separately, but did not implement the full transformer encoder. \textcolor{red}{\hl{}}Ye et al. \cite{ye_accelerating_2023} focused on accelerating only the attention mechanism using a reconfigurable systolic array for the transformer. Similarly, Zhang et al. \cite{zhang_algorithm-hardware_2021} concentrated on accelerating the attention layer through hardware-software co-design. In contrast, \textbf{\textit{ADAPTOR}} is developed to support the entire transformer neural network (TNN). Some other works accelerate the full transformer networks but their logic circuits go through the time-consuming synthesis steps for different models or they perform poorly on the same model with different configurations \cite{FaFlA}. These approaches lack the generality to support diverse variants, whereas \textbf{\textit{ADAPTOR}} eliminates the need for repeated synthesis across models.

As transformer variants continue to evolve with differing parameters, designing a generic and efficient accelerator that can be customized to the structural characteristics of these variants becomes increasingly valuable. Thus, a versatile accelerator is needed to efficiently handle dense matrix computations across various TNN applications, and \textbf{\textit{ADAPTOR}} is designed to fulfill this role. Digital signal processing (DSP) resources are capable of high-speed computation at higher frequencies. Proper utilization of them depends on the implementation method. For example, most accelerators \cite{peng_length_2022, peng_accelerating_2021, jiang_ultra_nodate, wojcicki_accelerating_2022} used high-level synthesis (HLS) tools, while some used hardware description language (HDL) \cite{chen_high-frequency_2023, yang_efa-trans_2022, bai_ltrans-opu_nodate} for design. 
While HLS requires less implementation time compared to HDL, writing efficient HLS code to use parallel DSPs for optimal performance is challenging \cite{lstm_high_rate}. To address this, \textbf{\textit{ADAPTOR}} employs optimized HLS coding techniques. Additional challenges include storing the vast number of TNN parameters in the on-chip memories of FPGAs, which typically have a size of 5MB for low-end devices such as the ZCU104 and 35MB for high-end devices such as the Alveo U200 \cite{qi_accelerating_2021} and executing the extensive number of multiplication and accumulation (MAC) operations required by TNNs on the DSPs, with Ultrascale+ FPGAs offering approximately 9024 DSPs. Therefore, input matrices must be partitioned into tiles. However, developing an optimal partitioning scheme that aligns well with the architecture presents a significant challenge, one that has been carefully addressed in the design of \textbf{\textit{ADAPTOR}}. The data access and computation patterns differ across various blocks within the transformer, which also prevents acceleration. To overcome this, \textbf{\textit{ADAPTOR}} assigns dedicated hardware modules to each block, enabling more effective design and optimization. The full source code \footnote{\url{https://github.com/Kabir-Ehsan/Transformer_on_FPGA}} to reproduce the presented results or improve the design. 

\hfill \\

In summary, this work makes the following contributions:
\begin{itemize}
    
    \item [$\bullet$] A novel accelerator architecture for a complete transformer that maximizes DSP and LUT utilization to enhance parallel processing and achieve low latency, supported by an analytical model for pre-execution estimates of resource use and latency. 
    
    \item [$\bullet$] An efficient tiling strategy for weight matrices in both the multi-head attention layer and the feedforward neural network layer, enabling the deployment of the accelerator to any FPGA platform for most TNN models.


    \item [$\bullet$] A modular design approach implemented using parameterized HLS codes to accommodate varying computation and data access patterns, as well as to allow design-time modification of different TNN components.
    

    \item [$\bullet$] A runtime adaptive feature allows software-driven parameter adjustments to run different models without hardware re-synthesis.

    
\end{itemize}

\section{Related Work}\label{relate}
Various custom and partially adaptive FPGA accelerators have been developed for TNNs. Peng et al. \cite{peng_length_2022, peng_accelerating_2021} introduced a coherent sequence length–adaptive algorithm-hardware co-design for Transformer acceleration and explored column-balanced block-wise pruning. Qi et al. \cite{qi_accommodating_2021, qi_accelerating_2021} proposed an acceleration framework combining balanced model compression at the algorithm level with hardware-level FPGA optimization. Chen et al. \cite{chen_understanding_2025} developed an analytical model for evaluating spatial TNN accelerators, considering FPGA compute and memory resources, identifying optimal parallelization and buffering strategies, and providing reusable HLS kernels. Similarly, we developed an analytical model to estimate the latency and resource utilization of \textbf{\textit{ADAPTOR}} and designed it modularly, with each module implemented as an HLS function for easy optimization and reuse. Qin et al. \cite{qin_enhance} designed a TNN accelerator with separate attention and linear kernels for long input sequences, applying tiling only to the attention layer, whereas our design applies unique tiling strategies to both attention and linear layers. Both architectures incorporate analytical models. The energy-efficient FTRANS framework \cite{li_ftrans_2020} employed an improved block-circulant matrix method for algorithm-level sparsity, alongside a dedicated accelerator designed for this approach. Most of these architectures target specific TNNs and sparsity patterns, lacking runtime flexibility to reconfigure the computing structure for different applications. In contrast, \textbf{\textit{ADAPTOR}} can be programmed from software for any dense TNN model. FlexRun \cite{fast} identified key NLP model components, implemented them on a state-of-the-art FPGA accelerator, performed design space exploration to determine the optimal architecture for a given NLP model, and enabled automatic reconfiguration based on the results. FET-OPU \cite{fet-opu} presented an overlay architecture for general TNN acceleration featuring a DSP-packed Matrix Multiplication Unit (MMU) with a FIFO-based data caching mechanism. FlightLLM \cite{flightLLM} introduced a configurable sparse DSP chain for handling diverse sparsity patterns efficiently, an always-on-chip decode scheme for improved memory bandwidth with mixed-precision support, and a length-adaptive compilation method to minimize instruction storage overhead for large language models. TRAC \cite{plagwitz_trac_2022} focused on dedicated hardware generation, integrating code generation into the compilation process to create parameterized and synthesized modules for specific Transformer configurations-unlike fixed, though parameterizable, overlays. EFA-Trans \cite{yang_efa-trans_2022} supports both dense and sparse computation patterns but requires hardware resynthesis to switch between them. Moreover, none of these works examined optimal tile sizes or DSP utilization for maximum parallelism as done in \textbf{\textit{ADAPTOR}}.

\section{Background}\label{back}
\subsection{Transformer Architecture}
\begin{figure}
\centering
    \begin{subfigure}{.5\textwidth}
    \centering
    \includegraphics[height=10cm, width=1.0\linewidth]{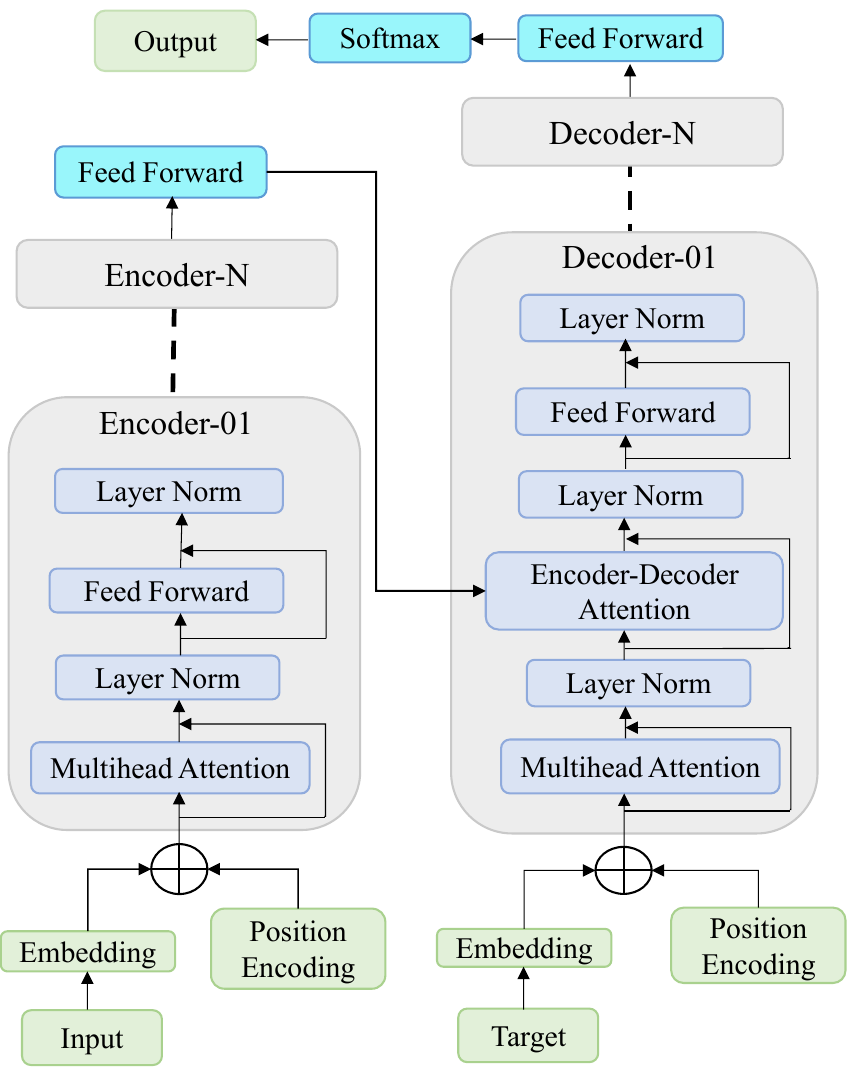}
    \caption{\label{TNN} Complete Architecture.}
    \end{subfigure}%
    \begin{subfigure}{.5\textwidth}
    \centering
    \captionsetup{justification=centering}
    \includegraphics[height=6cm, width=0.95\linewidth]{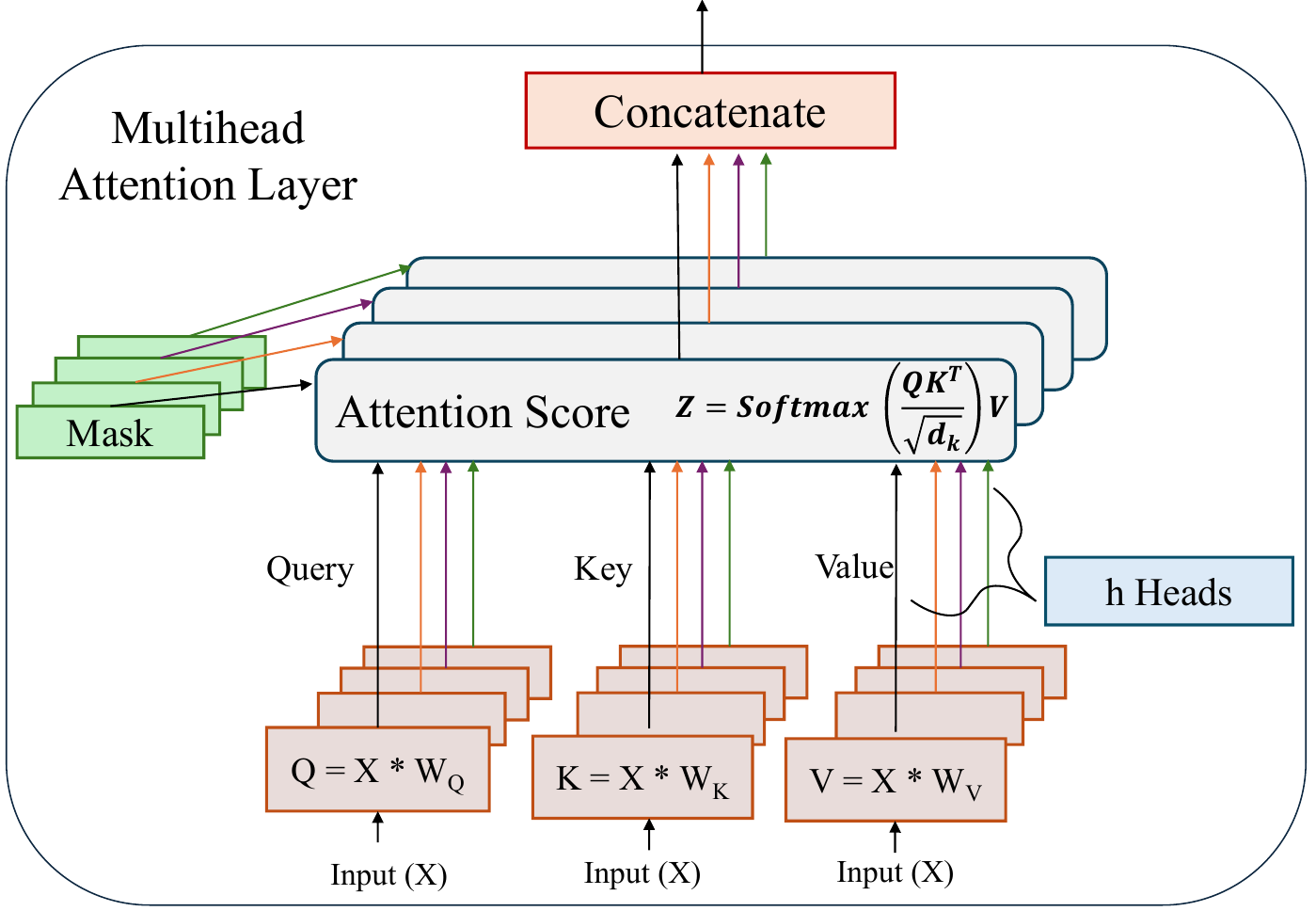}
    \hfill \\ 
    \hfill \\ 
    \hfill \\ 
    \hfill \\ 
    \hfill \\ 
    \hfill \\ 
    \caption{\label{mha}Multihead Attention Layer.}
    \end{subfigure}
\caption{Transformer Neural Network}
\end{figure}

There are several building blocks in transformers as shown in Fig.~\ref{TNN}. An input sequence of tokens is converted into embeddings. The positional encoder enables the model to consider the order of tokens in a sequence by adding positional information to the embeddings. It generates vectors that give context according to the word's position in a sentence. Then the vectors are linearly transformed into three tensors: Q (queries), K (keys), and V (values) by multiplying the embedding matrix with three weight matrices. The encoder block handles these tensors, transforming them into a higher-level representation that encapsulates crucial information. This process ensures the proper capture of features and contextual relationships within the input sequence. The encoder architecture comprises two main sub-layers: (1) the self-attention mechanism, and (2) the position-wise feed-forward network. The self-attention mechanism enables the model to assess different segments of an input sequence simultaneously. It captures long-range relationships by measuring attention scores and utilizing multi-head projections for various input representations. Thus, it can learn complex patterns, dependencies, and relationships effectively. The position-wise feed-forward network (FFN), which is equivalent to a multilayer perceptron (MLP), applies linear transformations to every position independently in the input sequence. In this network, two linear transformations are executed. They mainly contain matrix-vector multiplication. The first linear transformation has activation functions such as the Rectified Linear Unit (ReLU) or Gaussian Error Linear Unit (GeLU) but the second one does not have these. Furthermore, each sub-layer includes a residual connection combined with layer normalization (LN). This reduces \todo[disable]{does it solve it? Myabe reduces it?} the vanishing gradient problem during training. Residual addition and LN layers are inserted after each MHA and FFN. It mainly includes the addition of matrix elements and nonlinear functions.
The decoder block illustrated in Fig.~\ref{TNN} is responsible for generating the output sequence based on the encoded representations supplied by the encoder. Like the encoder, the decoder also consists of a stack of N identical layers. Each layer within the decoder contains three sub-layers. They are: (1) the Masked Attention Mechanism, resembling the encoder's self-attention, and it includes a masking feature that restricts the output's dependency on known preceding outputs; and (2) an attention layer that directs its focus to the encoder's output, enabling the decoder to emphasize relevant sections of the input sequence for each output element. and (3) a position-wise feed-forward network.

The self-attention mechanism in transformers allows each position in the sequence to attend to all other positions, enabling the model to consider global context easily. Each attention head is composed of three linear layers and a scaled dot-product attention function. The parameter $h$--or number of heads--is equal to 8 in the Transformer base model or 16 in the Transformer big model. As illustrated in Fig.~\ref{mha}, the scaled dot product attention in each head is a crucial part of the multihead attention layer. The attention weights are computed by performing the dot product of the query and key vectors and subsequently scaling it down by the square root of the dimension of the key vectors. This scaling is essential to prevent the dot products from becoming excessively large, which contributes to the stabilization of gradients during the training process. Subsequently, the scaled dot products undergo the softmax function, resulting in the computation of attention weights. These weights are then used to perform a weighted sum of the value vectors. The ultimate output is the projection of the concatenated sequences from all heads.

The output of MHA can be represented as Equation \ref{eq:attention} \& \ref{q_k_v}. The input sequence X is linearly mapped into $Q_i, K_i, V_i$ matrices using weights and biases. The parameter $d_k = d_{model}/h$ is the dimension of $Q_i$ and $K_i$. $d_{model}$ is a hyperparameter called embedding dimension, and $h$ is the number of heads. 
\begin{multicols}{2}   
\noindent\begin{gather}
\label{eq:attention}
\begin{split}
    Attention (Q_i, K_i, V_i) = \textcolor{red}{}softmax \left(\frac{Q_iK^T_i}{\sqrt{d_k}}\right)V_i   
\end{split}
\end{gather}
\begin{gather}
\label{q_k_v}
\begin{split}
    Q_i = X \times W_q + B_q, \\K_i = X \times W_k + B_k, \\V_i = X \times W_v + B_v
\end{split}
\end{gather}
\end{multicols}
\begin{multicols}{2}   
\noindent\begin{gather}
\label{ffn}
\begin{split}
    \textcolor{red}{}FFN(X) = &Layer\_Norm(X + \\ & ReLU(X \times W_1 + b_1) \times W_2 + b_2) 
\end{split}
\end{gather}
\begin{gather}
\label{ln_eq}
\begin{split}
Layer\_Norm(X) = \gamma \left( \frac{X-\mu}{\sqrt{{\sigma}^2 +\epsilon}} \right) + \beta
\end{split}
\end{gather}
\end{multicols}
\textcolor{red}{\hl{}}The FFN comprises a LN operation, residual addition, a ReLU activation, and two linear sublayers, as described in Equation \ref{ffn}, where W\textsubscript{1}, W\textsubscript{2} are weights and b\textsubscript{1}, b\textsubscript{2} are biases. The operations for layer normalization, softmax, GELU and RELU activation functions are described in equations \ref{ln_eq}, \ref{soft_eq}, \ref{gelu_eq}, and \ref{relu_eq} respectively, where X is the input vector (for a particular position in the sequence), $\mu$ is the mean of X, ${\sigma}^2$ is the variance of X, $\gamma$ and $\beta$ are learnable parameters, and $\epsilon$ is a small constant.
\begin{multicols}{2}
\noindent\begin{gather}
\label{soft_eq}
\begin{split}
softmax(X_j) = \frac{e^{X_j}}{\sum\limits_{i=1} e^{X_i}}
\end{split}\\
\label{gelu_eq}
\begin{split}
    GELU(x) = xP(X \le x) = x\times \frac{1}{2} [1 + erf(X/\sqrt(2))]
\end{split}
\end{gather}
\begin{equation}
\label{relu_eq}
    RELU(X) = 
        \begin{cases}
            0, & X < 0\\
            X, &  X \ge 0
        \end{cases}  
\end{equation}
\end{multicols}

\subsection{High Level Synthesis Design} 
High-Level Synthesis (HLS) allows designers to describe circuit functionality at a higher level of abstraction than that of hardware description language. HLS tools translate high-level code, typically written in languages like C, C++, or OpenCL, into Register-Transfer Level (RTL) code suitable for FPGA implementation. This approach offers several advantages, including faster development cycles and simplified design modifications, as designers can use familiar programming languages to describe the hardware. Moreover, HLS enables efficient design space exploration, allowing different architectures to be evaluated without extensive hardware design expertise, leading to the rapid creation of optimized accelerators optimized for power, performance, and area \cite{hlsScale}. However, HLS does come with challenges, such as ensuring that the generated RTL meets the specified constraints. The success of the synthesized hardware is largely dependent on the robustness of the HLS tools and the expertise of the designer.

\section{\textbf{\textit{ADAPTOR}}'s Architecture}\label{sec2} 
The core of the \textbf{\textit{ADAPTOR}} is designed in C language on Vitis high-level synthesis (HLS) 2022.2.1 tool. C simulation confirms the algorithm's correctness, while C/RTL co-simulation validates the functionality of the synthesized hardware. This section describes the HLS design technique that generates an optimized architecture utilizing most of the LUTs and DSPs in the processing modules, ensuring high parallelism of computation. 
There are loading units, computing modules, and activation function units in the overall architecture, which are described below. Fig.~\ref{mha_System} and \ref{FF_System} represent two main computing modules of \textbf{\textit{ADAPTOR}}.

\subsection{Attention Module}

\begin{figure}
\centering
\includegraphics[height=6cm, width=0.9\linewidth]{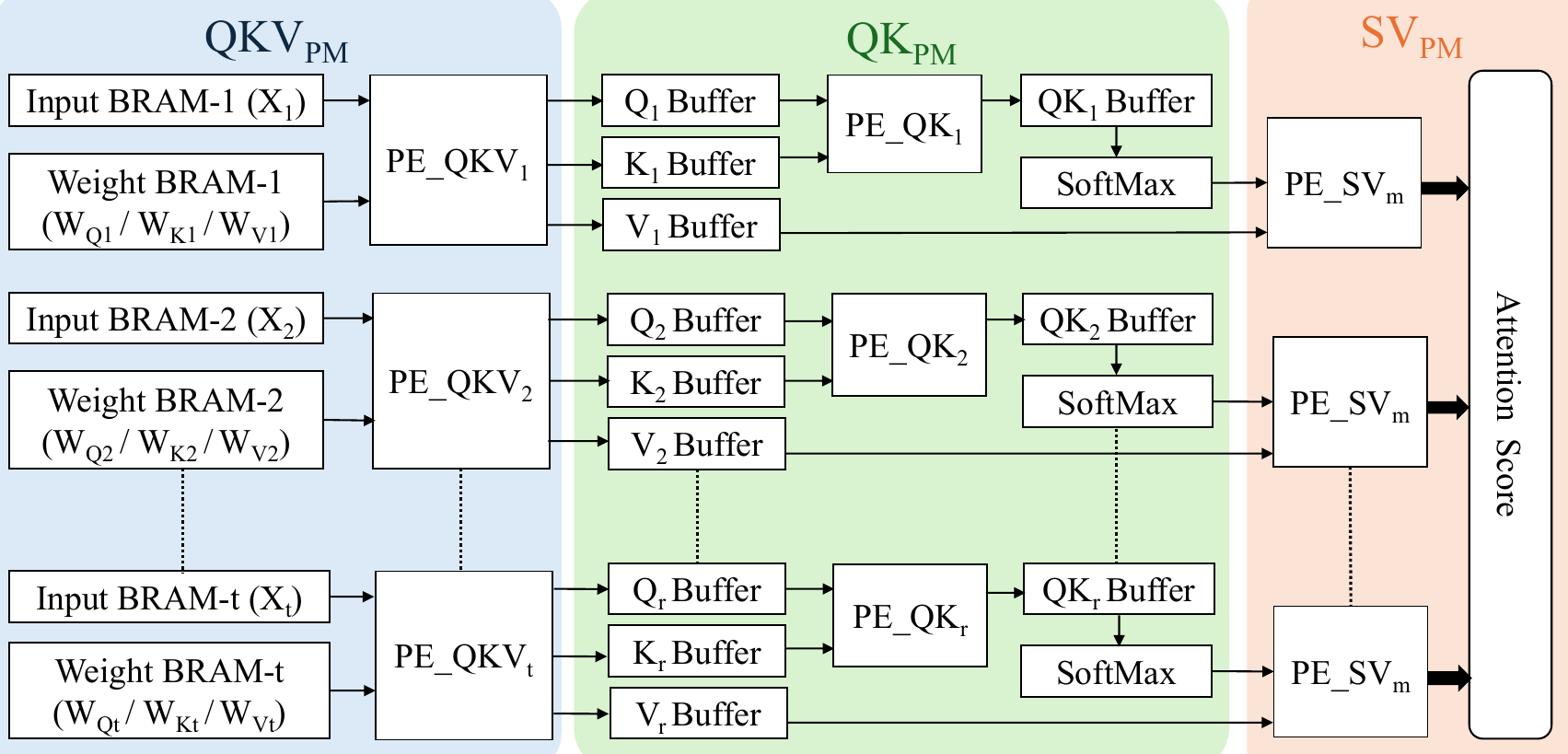} 
\caption{\label{mha_System}Attention Module of \textbf{\textit{ADAPTOR}}.}
\end{figure}

The overall architecture designed to accelerate the attention mechanism is illustrated in Fig.~\ref{mha_System}. It consists of three principal processing modules (PMs), denoted as ${QKV}_{PM}$, ${QK}_{PM}$, and ${SV}_{PM}$, according to the specific operations they perform. Each of these modules begins operation only after the previous module has completed its computations. This strict sequential execution ensures that all data dependencies are respected and simplifies control logic. The number of module instances corresponds to the number of attention heads ($h$). Within each module, computation is carried out by an array of processing elements (PEs), where each PE incorporates a DSP48 unit responsible for multiplication and accumulation (MAC) operations. The organization of the PE arrays varies across modules, as their computational demands and data access patterns differ. To accommodate these differences, the modules are implemented as separate functions in high-level synthesis (HLS), thereby enabling targeted optimization of the corresponding register-transfer level (RTL) components. Parallel data access is supported by distributing input activations and weights across multiple BRAMs and LUTRAMs.

Each PE operates independently, equipped with its own local memory, control logic, and computational resources. The weight matrices associated with the generation of queries ($W_q$), keys ($W_k$), and values ($W_v$) are stored as two-dimensional arrays of dimension $\left(\tfrac{d_{model}}{h} \times TS_{MHA}\right)$, where $TS_{MHA}$ denotes the tile size of the attention module. This tiling strategy partitions the larger weight matrices into sub-matrices, thereby facilitating efficient parallelization. The interplay between the number of heads, tiling parameters, and the HLS array partitioning directives determines how these arrays are mapped onto multiple two-port BRAMs. Since BRAM ports are limited, careful partitioning and scheduling of data transfers ensure that all operands required concurrently by the DSP units are accessible without contention. The intermediate $Q$, $K$, and $V$ matrices, each of size $\left(SL \times \tfrac{d_{model}}{h}\right)$ where $SL$ denotes the sequence length, are buffered locally to support subsequent stages of computation.

\subsubsection{\textbf{{QKV}\textsubscript{PM} module:}}

The ${QKV}_{PM}$ module is responsible for generating the query, key, and value matrices. It incorporates dedicated BRAMs for the weights ($W_Q$, $W_K$, $W_V$) and for the input activations ($X_i$), which provide parallel data access to the DSP units within the processing element (PE) array. To accommodate on-chip memory constraints, the weight and input arrays are divided into subarrays using a tiling strategy, ensuring that the data can be efficiently mapped onto BRAMs or LUTRAMs. The number of times the ${QKV}_{PM}$ module is invoked is determined by the tiling factor, resulting in a total of $\tfrac{d_{model}}{TS_{MHA}}$ iterations. At each iteration, the buffers for $W_Q$, $W_K$, $W_V$, and $X_i$ are populated with distinct tiles of data, after which computation is initiated within the PEs.

During these operations, the corresponding bias terms for the $Q$, $K$, and $V$ matrices are fetched from off-chip memory into registers in parallel with the primary computations of the ${QKV}_{PM}$ module. These biases are subsequently integrated into the generated matrices, thereby completing the linear transformations. The computational flow of this module is summarized in Algorithm~\ref{QKV} of Section~\ref{sm}, where pipelining of the outer loop facilitates full unrolling of the innermost loop. This design yields an array of $\tfrac{d{model}}{TS_{MHA}}$ PEs, thereby maximizing throughput while maintaining an efficient mapping of resources.

\subsubsection{\textbf{{QK}\textsubscript{PM} module:}}

The ${QK}_{PM}$ module carries out the matrix–matrix multiplication between the $Q$ and $K$ matrices. Since these matrices are relatively small in dimension, tiling is not required. The computational flow is summarized in Algorithm~\ref{QK} of Section~\ref{sm}, where full unrolling of the innermost loop produces $\tfrac{d{model}}{h}$ processing elements (PEs). Within this module, the $Q$ and $K$ matrices are buffered to enable parallel access by the DSP units. In addition to the multiplication operations, the division specified in Equation~\ref{eq:attention} is also performed within this module using LUT resources. To avoid excessive LUT utilization, the degree of parallelism for this operation is deliberately constrained. The output of this module is the intermediate attention weight matrix $S$, which is stored in either BRAMs or registers depending on availability and access requirements. These weights are subsequently passed to the non-linear softmax function, implemented in HLS using LUTs and flip-flops, to complete the attention score computation.

\subsubsection{\textbf{{SV}\textsubscript{PM} module:}}
The normalized attention weight matrix ($S$), obtained from the softmax operation, is supplied to the ${SV}_{PM}$ module, where it is combined with the value ($V$) matrix through matrix–matrix multiplication. As described in Algorithm~\ref{SV} of Section~\ref{sm}, the innermost loop is fully unrolled, enabling $SL$ processing elements to operate in parallel. The resulting output, referred to as the attention score, represents a weighted aggregation of the value vectors and constitutes the final contribution of the attention mechanism to the subsequent layers.
\subsection{Feedforward Network Module}
\begin{figure}
\centering
\includegraphics[height=11cm, width=0.9\linewidth]{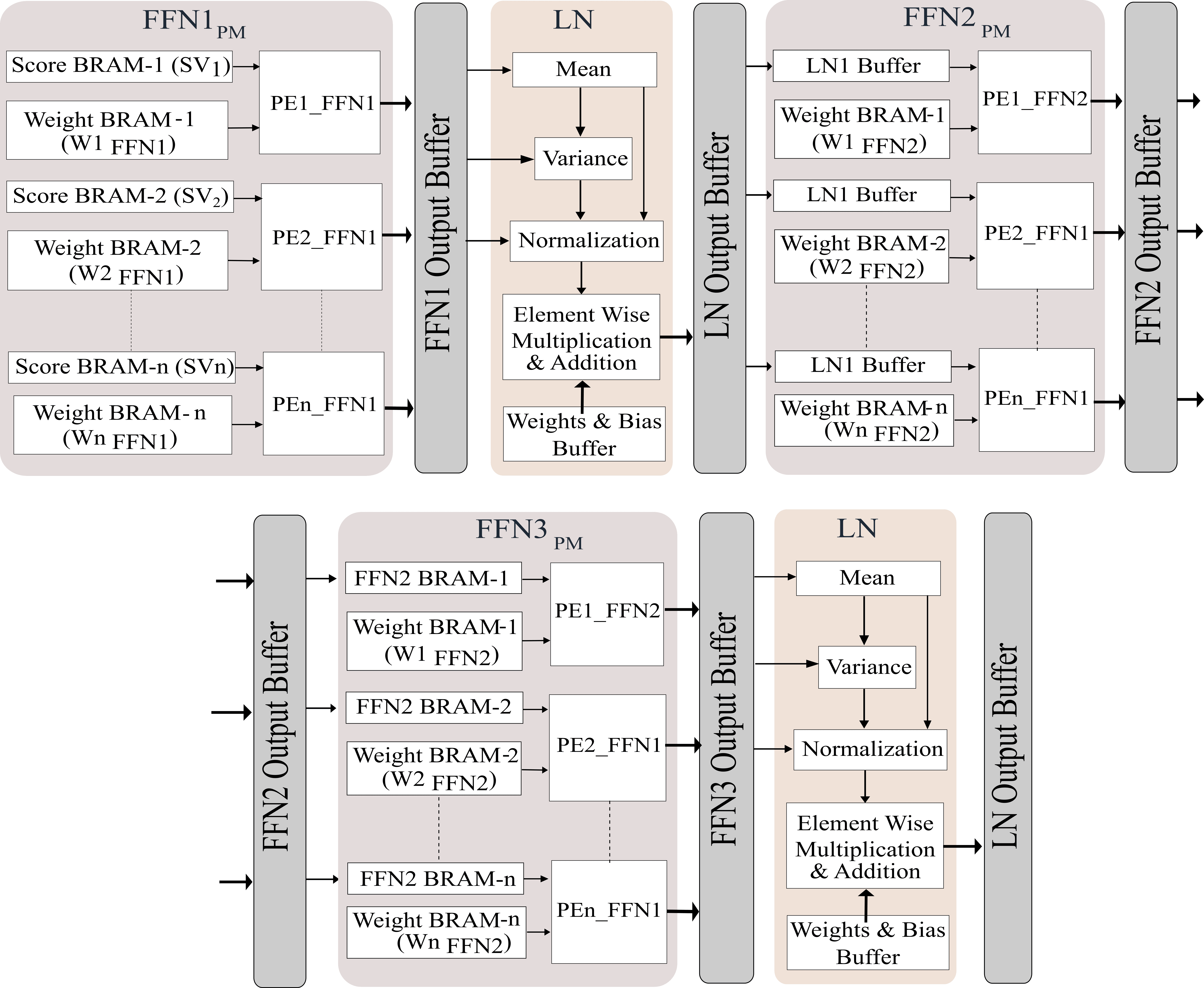} 
\caption{\label{FF_System}Feedforward Network Module of \textbf{\textit{ADAPTOR}}.}
\end{figure}

The architecture developed to accelerate the feedforward network (FFN) module is depicted in Fig.~\ref{FF_System}. Three distinct RTL modules $FFN1_{PM}$, $FFN2_{PM}$, and $FFN3_{PM}$ are implemented to support variations of the FFN across different architectural configurations. They are executed sequentially to respect the inherent data dependencies. Each module begins processing only after the preceding module has fully completed its computation. In high-level synthesis (HLS), these modules are described as separate functions, each defined by input and output arrays of different dimensions, which are subsequently mapped onto BRAMs or LUTRAMs during synthesis. Since the computational workload differs across the modules, each function is optimized independently, resulting in varying numbers of processing elements depending on the unrolling factor applied to the innermost loop. The weights of the FFN are stored in a two-dimensional array ($W_o$) of dimensions $\left(\tfrac{d_{model}}{TS_{FFN}} \times \tfrac{4\times d_{model}}{TS_{FFN}}\right)$, where $TS_{FFN}$ denotes the tile size in FFN. This tiling strategy partitions the weight matrices into smaller blocks, facilitating parallel access and efficient memory utilization. Among the three RTL modules, both $FFN1_{PM}$ and $FFN3_{PM}$ are followed by layer normalization (LN), ensuring stabilized activations before passing results to subsequent stages.

\subsubsection{\textbf{$FFN1$\textsubscript{PM} module}}
The $FFN1_{PM}$ module performs the initial linear transformation on the attention scores, serving as the first stage of the feedforward network. To accommodate on-chip memory constraints, the arrays used by the processing elements (PEs) are tiled along both dimensions. Consequently, the module is invoked $TS_{FFN} \times TS_{FFN}$ times to complete the transformation. As outlined in Algorithm~\ref{FFN1_alg} of Section~\ref{sm}, pipelining of the second loop enables full unrolling of the innermost loop (line 7), producing $TS_{FFN}$ PEs in total. This corresponds to $\tfrac{d_{model}}{\text{No.\ of Tiles}\_{FFN}}$ parallel computational units.

\subsubsection{\textbf{$FFN2$\textsubscript{PM} module}}
Building upon the normalized outputs of $FFN1_{PM}$, the $FFN2_{PM}$ module performs the second linear transformation, expanding the intermediate representation. Similar to $FFN1_{PM}$, arrays are tiled along both dimensions, though this module requires $4 \times TS_{FFN} \times TS_{FFN}$ accesses due to the increased dimensionality of the operation. The computational flow is summarized in Algorithm~\ref{FFN2_alg} of Section~\ref{sm}, where pipelining again enables full unrolling of the innermost loop (line 7). This results in $TS_{FFN}$ PEs, corresponding to $\tfrac{d_{model}}{\text{No.\ of Tiles}\_{FFN}}$ units of parallelism, consistent with the structural design of the first module.

\subsubsection{\textbf{$FFN3$\textsubscript{PM} module}}
The $FFN3_{PM}$ module applies the final linear transformation to the normalized outputs of $FFN2_{PM}$, projecting them back to the original model dimension. As with the preceding modules, arrays are tiled along both dimensions, requiring $4 \times TS_{FFN} \times TS_{FFN}$ iterations to complete the computation. Algorithm~\ref{FFN3_alg} of Section~\ref{sm} describes the workflow, where pipelining and full loop unrolling (line 7) yield $4 \times TS_{FFN}$ PEs. This corresponds to $\tfrac{4 \times d_{model}}{\text{No.\ of Tiles}\_{FFN}}$, reflecting the higher dimensionality of this stage.

\subsection{Load Weights Unit}
Three dedicated \textit{Load\_{Weights}} units are employed to manage the transfer of parameters from external memory to on-chip buffers. The first unit supplies weights to the weight memories of the attention heads (Fig.~\ref{mha_System}), while the second serves the feedforward network (Fig.~\ref{FF_System}). A third unit is responsible for loading the weights associated with the layer normalization modules. This separation ensures that weight data can be delivered efficiently to each functional block in accordance with its computational demands. For the attention module, weights are represented in HLS as two-dimensional arrays of dimension $\left(\tfrac{d_{model}}{h} \times TS_{MHA}\right)$, where $TS_{MHA}$ denotes the tile size applied to partition the larger matrices into sub-matrices. After synthesis, these arrays are mapped onto dual-port BRAMs or LUTRAMs, and are populated iteratively with tile-specific data transferred from external memory at each iteration. In the feedforward network, the weight arrays are defined with dimensions $\left(TS_{FFN} \times 4\times TS_{FFN}\right)$. Here, $TS_{FFN}$ corresponds to the tiling parameter, defined as $\tfrac{\text{Embedding Dimension}}{\text{No.\ of Tiles}\_{FFN}}$, while $4\times TS_{FFN}$ equals $\tfrac{\text{Hidden Dimension}}{\text{No.\ of Tiles}\_{FFN}}$. These weights are therefore partitioned along both row and column dimensions, requiring iterative loading across tiles. As in the attention module, they are synthesized as dual-port BRAMs or LUTRAMs. The weights for layer normalization are comparatively simple, represented as one-dimensional arrays of length $d_{model}$. As no tiling is required in this case, the entire weight set is transferred in a single step and subsequently synthesized into dual-port BRAMs or LUTRAMs. The complete procedure for weight loading is outlined in Algorithm~\ref{loadwq} of Section~\ref{sm}.

\subsection{Load Inputs Unit}
Input data are transferred from external memory into dedicated input BRAMs, which are implemented in HLS as dual-port, two-dimensional arrays of size $\left(SL \times d_{model}\right)$, where $SL$ denotes the sequence length. These BRAMs are reused across encoder and decoder layers, allowing each layer to access the outputs of the previous layer as inputs for subsequent computations. Three distinct \textit{Load\_inputs} units manage the data movement to accommodate differences in computation, tiling, and array dimensions across modules. The first unit populates the intermediate input BRAMs of each attention head (Fig.~\ref{mha_System}) using Algorithm~\ref{loadin_MHA} of Section~\ref{sm}. These BRAMs are represented as two-dimensional arrays of size $\left(SL \times TS_{MHA}\right)$, where tiling is applied along the column dimension. Consequently, data are loaded iteratively $\tfrac{d_{model}}{TS_{MHA}}$ times to supply all columns to the processing elements.

The second unit transfers data to the Score BRAMs of the $FFN1_{PM}$ module (Fig.~\ref{FF_System}), defined as two-dimensional arrays of size $\left(SL \times TS_{FFN}\right)$, using Algorithm~\ref{loadin_FFN1}. The third unit supplies data to the LN1 buffers of the $FFN2_{PM}$ module (Fig.~\ref{FF_System}), represented as arrays of size $\left(SL \times 4 \times TS_{FFN}\right)$ and loaded according to Algorithm~\ref{loadin_FFN3} of Section~\ref{sm}. The separation of load units ensures efficient handling of the differing computational demands, tile sizes, and array shapes across the attention and feedforward network modules.

\subsection{Load Biases Unit}
Bias parameters are stored in registers due to their relatively small size, enabling low-latency access during computation. Three dedicated \textit{Load\_bias} units manage the transfer of biases from external memory to the corresponding registers. The first unit supplies biases to the registers of each attention head in accordance with Algorithm~\ref{loadbiasMHA}. The same procedure, as described in Algorithm~\ref{loadbiasAll} of Section~\ref{sm}, is used to load biases for the feedforward network and the layer normalization modules. In HLS, biases are represented as one-dimensional arrays, and the application of a complete array partition pragma maps these arrays directly to registers. Since tiling is unnecessary for these small vectors, each array is loaded in a single transfer, providing all bias values simultaneously.
\subsection{Activation Unit}
The activation functions employed within the transformer architecture are implemented at this stage. Commonly used functions include ReLU, GeLU, and softmax, each defined according to its mathematical formulation. After synthesis, these functions are realized using LUTs to support efficient hardware computation. While the implementations of ReLU and GeLU are straightforward, the softmax function involves more complex operations; therefore, only its implementation is detailed in Algorithm~\ref{soft_alg} of Section~\ref{sm}.
\subsection{Layer Normalization Unit}
The Layer Normalization (LN) unit computes the mean and variance of the outputs from both the attention and feedforward network layers, following Equation~\ref{ln_eq}. These statistics are then used to normalize the outputs, which are subsequently scaled by learned weights and shifted by biases in an element-wise manner. Notably, the outputs of $FFN1_{PM}$ and $FFN3_{PM}$ are processed through the LN unit prior to subsequent stages, as illustrated in Fig.~\ref{FF_System}. The corresponding HLS implementation is provided in Algorithm~\ref{LNorm}  of Section~\ref{sm}.

\subsection{Bias Add Unit}
Three dedicated \textit{Bias\_add} units are employed to incorporate biases into the query ($Q$), key ($K$), and value ($V$) matrices, as well as into the outputs of the three feedforward network modules. Separate units are necessary because the corresponding HLS functions operate on arrays of differing dimensions, producing outputs with distinct shapes. One of the units associated with the feedforward networks additionally integrates the ReLU activation function. The operations performed by these units are detailed in Algorithms~\ref{ba1_alg}, \ref{ba2_alg}, and \ref{ba3_alg} of Section~\ref{sm}.

\section{System Design and Optimizations}\label{system}
\begin{figure}
\centering
\includegraphics[height=5.0cm, width=0.7\linewidth]{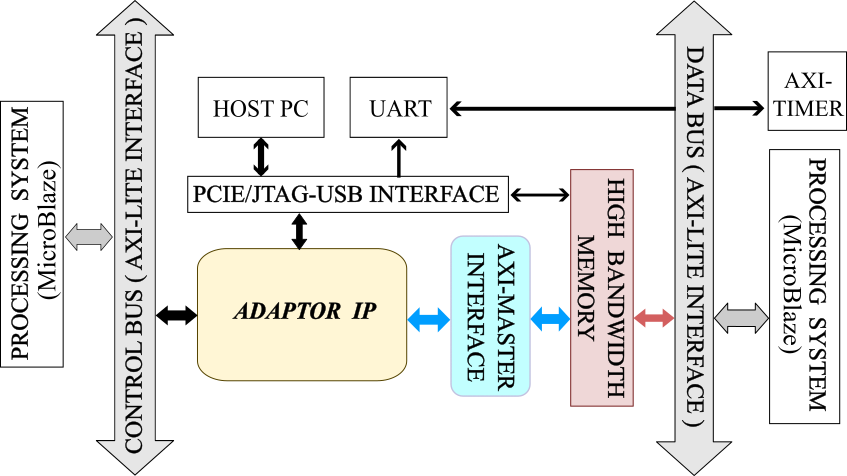}
\caption{\label{System}Complete System Design}
\end{figure}

\emergencystretch 3em
The overall system design for deploying \textbf{\textit{ADAPTOR}} across multiple FPGA platforms is presented in Fig.~\ref{System}. Experiments were conducted on three representative devices: the VC707 (Virtex-7 xc7vx485tffg1761-2), ZCU102 (Zynq UltraScale+ xczu9eg-ffvb1156-2-e MPSoC), and the Alveo U55C (UltraScale+ xcu55c-fsvh2892-2L-e). While the VC707 and ZCU102 boards integrate on-board DDR3 DRAM, the Alveo U55C employs high-bandwidth memory (HBM), offering significantly greater throughput for memory-intensive workloads.

Design parameters such as the number of attention heads, embedding dimension, hidden dimension, sequence length, and the number of encoder and decoder layers can be reconfigured at runtime, up to their maximum supported values, via a MicroBlaze softcore processor using the AXI4-Lite interface. These parameters are stored in a set of registers within \textbf{\textit{ADAPTOR}} to specify the topology of the TNN during runtime from the software. The registers are described in Table~\ref{config_regs}, along with the corresponding parameters they store. The system architecture was implemented using the Vivado 2022.1.2 design suite. The accelerator itself is encapsulated in a custom IP block generated from high-level synthesis (HLS) and integrated into the larger system. All toolflows from Xilinx-AMD were executed on a host workstation equipped with an Intel(R) Xeon(R) Gold 6130 CPU (2.10 GHz, 32 cores) and 192 GB of RAM. Data movement between the accelerator and external memory is managed through AXI4 master interfaces \cite{axiMaster}. Depending on workload demand, the accelerator fetches inputs and weights directly from off-chip HBM or DRAM when instructed by the accelerator controller. Control signals are delivered from the processor to the accelerator via an AXI-Lite slave interface. Additionally, the processor facilitates data transfers between external memory and on-chip BRAMs, while also issuing configuration and synchronization signals to the accelerator. 

The boards were connected to the host system through either a USB–JTAG interface or PCIe 3.0 $\times$4 link. Although the system includes a DMA/Bridge Subsystem for PCIe IP \cite{DMA}, PCIe-based communication was not utilized in this work. Performance measurements were conducted using the AXI-TIMER \cite{axitimer}, which recorded end-to-end latency between the initiation and completion signals of the custom IP. Final results were communicated back to the host via the UARTLite interface \cite{axiuart}, with outputs displayed on the terminal connected through the JTAG interface \cite{jtag}.

\begin{table}[h!]
\centering
\caption{Configuration Registers of the \textit{ADAPTOR}}
\label{config_regs}
\begin{tabular}{|c|c|}
\hline
\textbf{Register Name} & \textbf{Description / Stored Parameter} \\ \hline
\texttt{Sequence} & Sequence length of inputs \\ \hline
\texttt{Heads} & Number of attention heads \\ \hline
\texttt{Layers\_enc} & Number of encoders \\ \hline
\texttt{Layers\_dec} & Number of decoders \\ \hline
\texttt{Embeddings} & Dimension of the embedding layer \\ \hline
\texttt{Hidden} & Dimension of the intermediate layers \\ \hline
\texttt{Out} & Number of outputs \\ \hline
\end{tabular}
\end{table}

\begin{figure}
\centering
\includegraphics[height=5cm, width=0.9\linewidth]{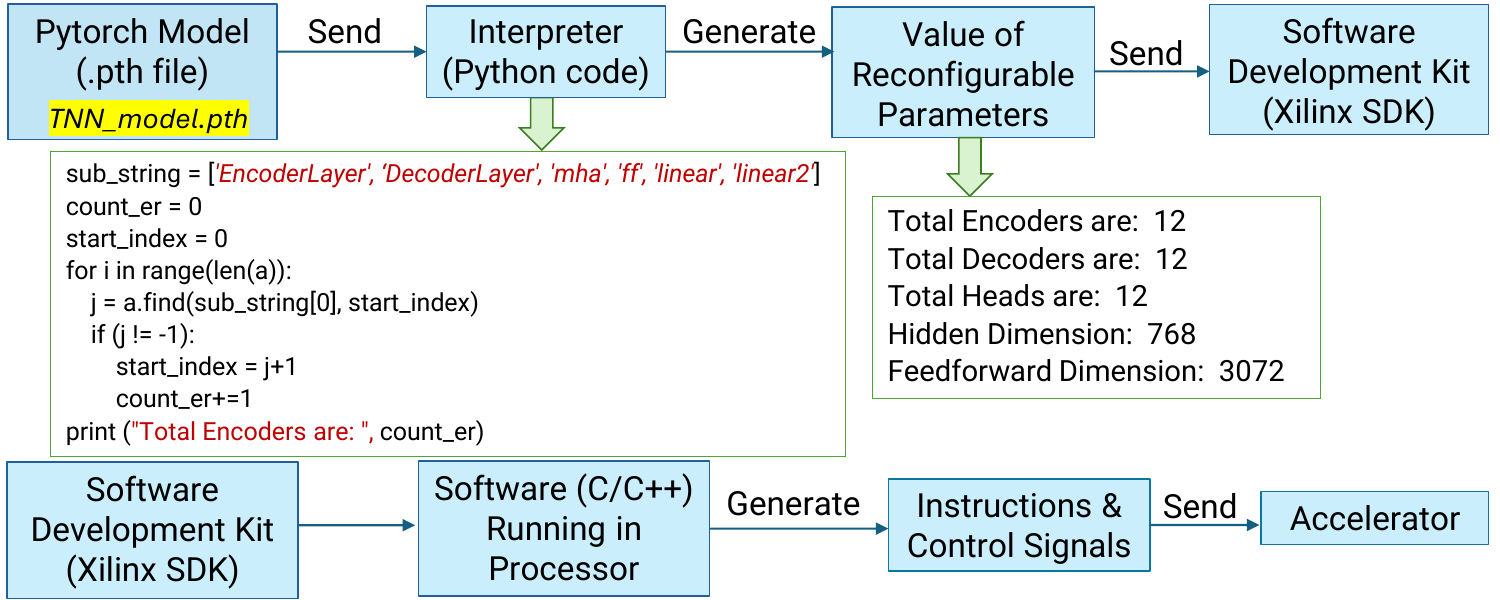} 
\caption{\label{isa} Programming procedures with software.} 
\end{figure}

The software interface illustrated in Fig.~\ref{isa} is designed to communicate the programmable parameters mentioned above to the accelerator. To support this, TNN models are trained using the PyTorch framework, with the trained models stored as \textit{'.pth'} files. In our experiments, we utilized publicly available pre-trained models from Hugging Face \cite{huggingfaceBERT}, trained on a Tesla V100 GPU. The software stack processes these files through a Python interpreter, which extracts the relevant parameter values. While these values vary across applications, the accelerator itself does not require re-synthesis for each case. Instead, only a subset of variables within the software must be reassigned to reflect the new configuration. The software, implemented in C++ using the Xilinx SDK and executed on the embedded processor, is summarized in Algorithm~\ref{sp} of Section~\ref{sm}. Based on the extracted parameters, the processor generates the necessary instructions and control signals to configure the accelerator, thereby enabling selective activation of different hardware components.

The software control overhead varies depending on the execution environment and task complexity. Loading a PyTorch model in Jupyter Notebook typically takes less than 2 seconds for small models and 5 to 20 seconds for large transformer models. Running a Python script, which primarily generates the C code executed in Vitis IDE, generally completes within a few seconds, depending on script complexity and hardware resources. In Vitis IDE, compiling and running the generated C program on the ARM processor (with the bitstream already programmed) requires about 5 to 30 seconds for compilation and 1 to 5 seconds to download and execute the ELF file; rerunning a prebuilt ELF takes less than 2 seconds before execution begins. The FPGA hardware kernels, already synthesized in the bitstream, execute directly with negligible control overhead, and their performance has been reported and compared in the results section.

\begin{figure}
\centering
    \begin{subfigure}{.5\textwidth}
    \centering
    \includegraphics[height=9cm, width=0.9\linewidth]{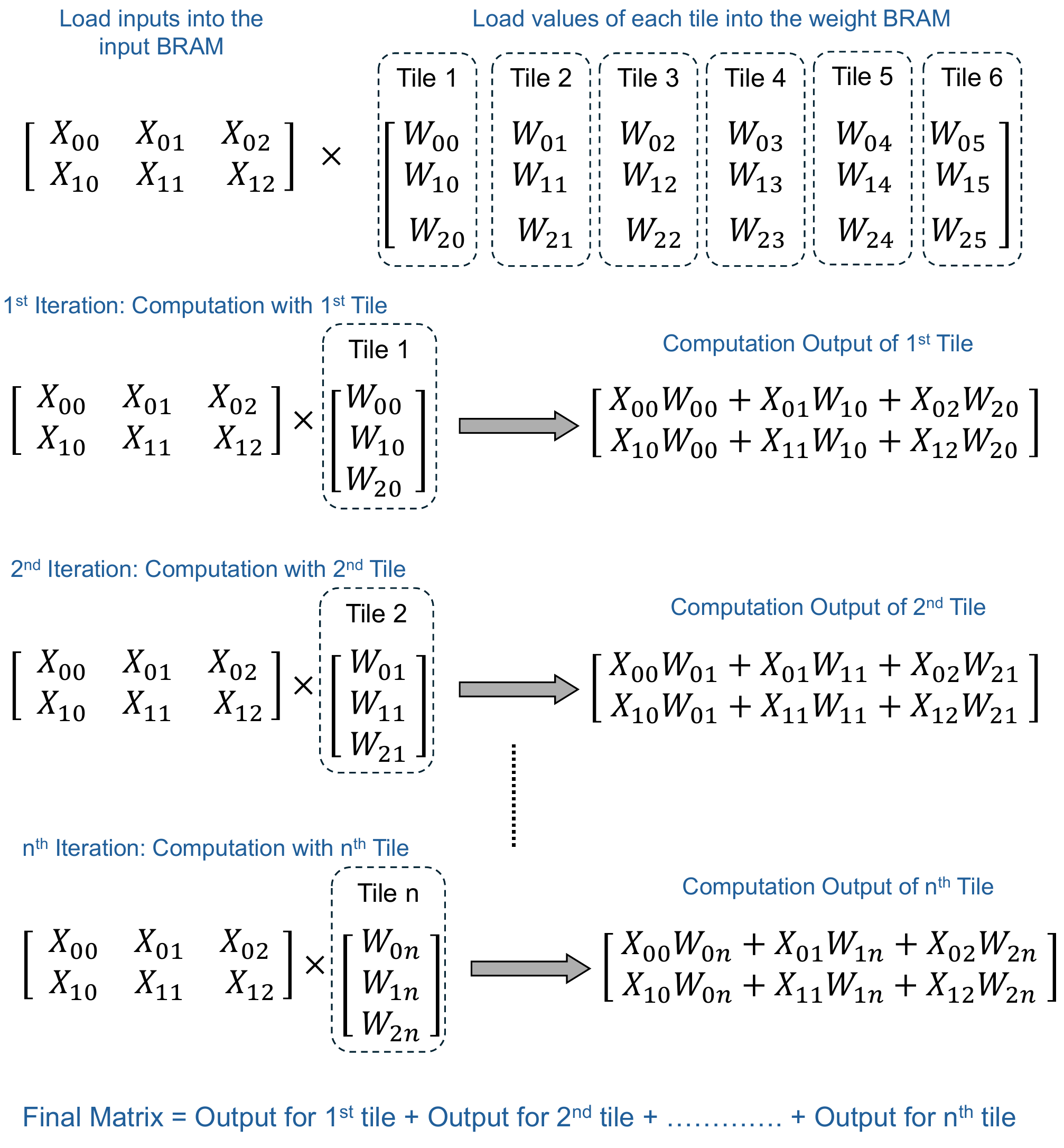}
    \caption{\label{tile_tech_mha} Tiling Technique in MHA.}
    \end{subfigure}%
    \begin{subfigure}{.5\textwidth}
    \centering
    \captionsetup{justification=centering}
    \includegraphics[height=7cm, width=1.0\linewidth]{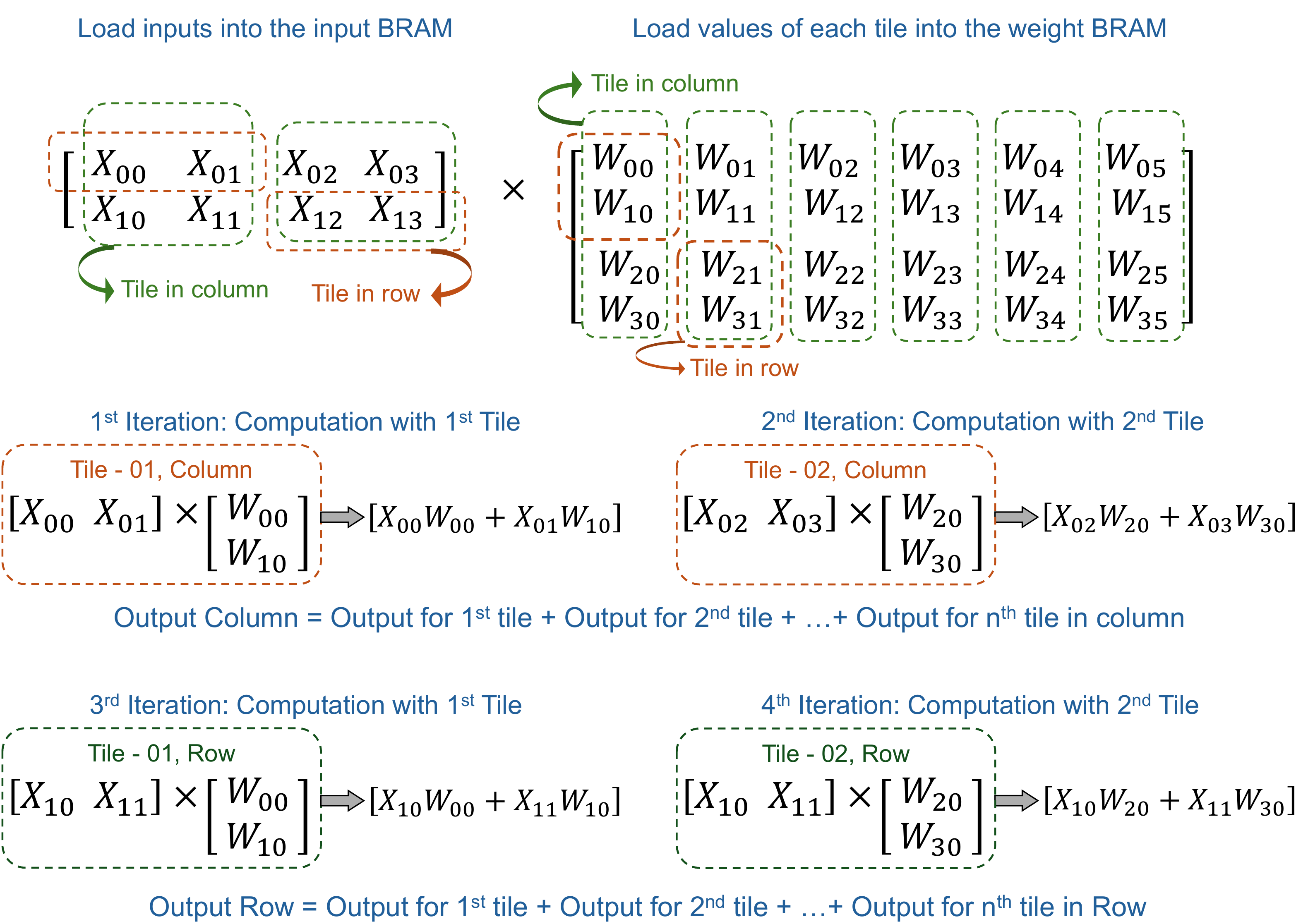}
    \hfill \\ 
    \hfill \\ 
    \hfill \\ 
    \caption{\label{tile_tech_FFN}Tiling Technique in FFN.}
    \end{subfigure}
\caption{Tiling Technique.}
\end{figure}

Transformer models are inherently large, leading to significant demands on both on-chip memory and computational resources. To address these challenges, we adopt a tiling strategy that enables efficient utilization of available hardware resources while maintaining manageable compilation times. Tiling facilitates the effective partitioning of arrays by the HLS tool, which in turn allows loop pipelining and unrolling to reduce computational latency. The proposed tiling approach for multi-head attention (MHA) is illustrated in Fig.~\ref{tile_tech_mha}.

In the attention module, the weight matrices are partitioned into tiles, enabling partial data loading from off-chip memory into BRAMs. Tiling is applied along the second dimension (the columns of the matrix), since the first dimension (the rows) is already reduced by the number of attention heads. Consequently, the weight matrices are loaded $\frac{d_{model}}{{TS}_{MHA}}$ times. Similarly, the input buffers for each attention head are defined as two-dimensional arrays of size $(SL \times TS_{MHA})$, and tiling is applied along the column dimension. The buffers are replenished $\frac{d_{model}}{{TS}_{MHA}}$ times, with one tile being processed at each iteration. During each iteration, the PEs compute results on the loaded tile, store intermediate results in buffers, and accumulate these with the outputs from previous iterations. The final output is thus obtained as the cumulative sum across all tiles. The feedforward networks (FFNs) following the attention layer are the most computationally demanding components of the encoder. Their weight matrices are represented as two-dimensional arrays of size $(TS_{FFN}) \times (4 \times TS_{FFN})$ and are tiled along both dimensions (rows and columns). Iterative loading is performed using two nested loops, one for each tiling dimension. As a result, the first FFN module is reused $(\frac{d_{model}}{TS_{FFN}})^2$ times, since both loops iterate $\frac{d_{model}}{TS_{FFN}}$ times. The second and third FFN modules are reused $(\frac{4 \times (d_{model})^2}{(TS_{FFN})^2})$ times due to their larger dimensions. The tiling strategy for the FFN is summarized in Fig.~\ref{tile_tech_FFN}, where intermediate results are first accumulated along the columns and subsequently along the rows to produce the final outputs across all tiles.

\begin{figure}
\centering
\includegraphics[height=6cm, width=1.0\linewidth]{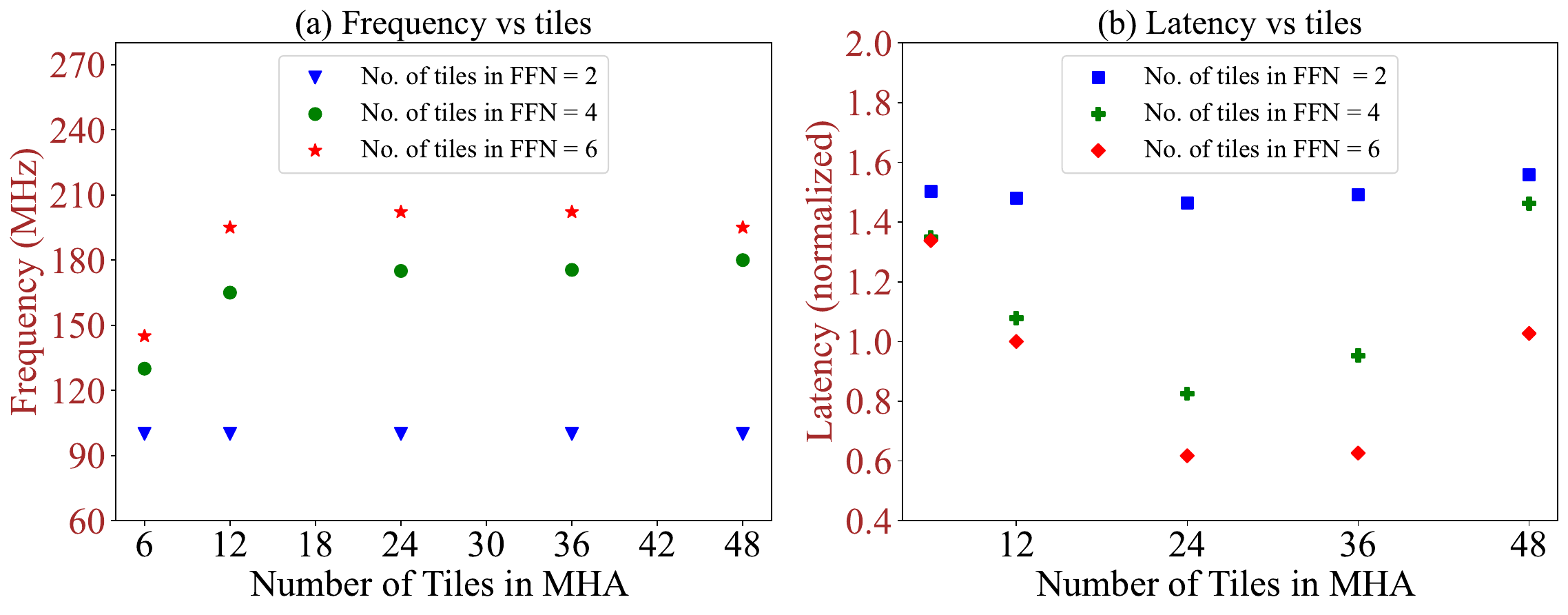} 
\caption{\label{tile_select}Choosing the optimum tile size.}
\end{figure}

The tile size must be fixed before synthesis, since changing it would require re-synthesizing the hardware. Figure~\ref{tile_select}(a) and (b) show how different choices of $TS_{MHA}$ and $TS_{FFN}$ affect both system frequency (MHz) and latency (normalized to the minimum value). In these experiments, the number of tiles in MHA ($\frac{d_{model}}{TS_{MHA}}$) was varied between 6 and 48, while the FFN tile count ($\frac{d_{model}}{TS_{FFN}}$) ranged from 2 to 6. The results highlight that using 24 tiles for MHA together with 6 tiles for FFN yields the best overall performance, reaching the highest frequency of 200 MHz and the lowest latency.

\section{Theoretical Model}\label{analyt}
The primary parameters influencing both resource utilization and performance in \textbf{\textit{ADAPTOR}} include the tile size, or equivalently the number of tiles, in the attention module and the feedforward network, as well as the number of attention heads, sequence length, embedding dimension, hidden dimension, and the number of encoder and decoder layers, assuming a fixed bit width. The utilization of DSPs is largely determined by the degree of parallelism in multiplication operations, with the highest demand observed in the ${QKV}_{PM}$, ${QK}_{PM}$, ${SV}_{PM}$, and $FFN$ modules. In contrast, BRAM utilization depends on the number of arrays required for intermediate data storage, the synthesis modes assigned to these memories, and the partitioning strategies specified through HLS pragmas.

To guide design-space exploration, we developed an analytical model that captures the relationship between these architectural parameters and the resulting latency and resource consumption. This model enables designers to predict performance and utilization outcomes, thereby facilitating informed parameter selection prior to full hardware synthesis.

\subsection{Model for DSP utilization}
Equation~\ref{eq:dsp} gives an estimate for DSP consumption. It was derived from all the loops described in the functions that generate RTL modules for ${QKV}_{PM}$, ${QK}_{PM}$, ${SV}_{PM}$, and $FFN$.
\begin{equation}
\label{eq:dsp}
\begin{split}
    No.\ of\ DSPs &= 3\times h \times \frac{d_{model}}{Tile\ no.\ MHA} + h\ \times \left(\frac{d_{model}} 
    {h}+SL\right) 
           + 6\times \frac{d_{model}}{Tile\ no.\ FFN}\ + d_{model} 
\end{split} 
\end{equation} 


The design follows a modular approach, with each module implemented as a function containing loops. The latency of a module depends on the time taken by its loops, which is affected by loop pipelining and unrolling directives. For nested loops, the second-to-last loop is pipelined, while the innermost loop is fully unrolled. The outermost loop is not modified with pragmas to avoid overly complex pipeline depth and high resource usage. The latency of a pipelined loop (PLL) can be calculated using Eq. \ref{LL}. If a pipelined loop is inside another loop, the total latency (TL) is given by Eq. \ref{TLL}. Here, the loop trip count (TC) is the number of iterations, and the initiation interval (II) is the time between the start of two consecutive iterations. Pipeline depth is the time needed to complete one iteration, depending on the sequential and parallel operations within it. Different modules can have different pipeline depths (PD). Latency is measured in clock cycles (cc).

\begin{gather}
\label{LL}
\begin{split}
   Pipelined\_Loop\_Latency = Pipeline\_Depth + Initiation\_Interval\times(Trip\_Count - 1) \ \ \text{\cite{perform}}
\end{split}\\
\label{TLL}
\begin{split}
   Total\_Latency = Pipelined\_Loop\_Latency\times Outer\_Loop\_Trip\_Count \ \ \text{\cite{perform}}
\end{split}
\end{gather} 

Equation \ref{LL} \& \ref{TLL} are generalized equations for measuring latency, the variables of which differ for different modules of \textbf{\textit{ADAPTOR}} as shown in the following equations.  
\subsection{Latency model for Attention Module}
\begin{multicols}{2}
\noindent\begin{gather}
\label{LI}
\begin{split}
  LI = [(d_{model}-1)\times 1 + PD\_L]\times SL 
\end{split}\\
\label{LB}
\begin{split}
  LBA = (\frac{d_{model}}{h}-1)\times 1 + PD\_L 
\end{split}
\end{gather}
\begin{gather}
\label{LWA}
\begin{split}
      LWA = [(\frac{d_{model}}{h}-1) \times 1 + PD\_L] \times SL 
\end{split}\\
\label{LIA}
\begin{split}
   LIA =& [(\frac{d\_{model}}{Tile\ no.\ MHA}-1) \times 1 + PD\_L] \times SL
\end{split}
\end{gather}
\end{multicols}

where, PD\_L is $Pipeline\_Depth\_Load$ that includes the time required to establish communication with HBM using AXI master interface (7 cc), read address location (1 cc), load (1 cc), and store (1 cc) data from and to that address, and convert floating point data to fixed point (3 cc) for tasks such as loading all inputs (LI), as well as loading inputs (LIA), biases (LBA) and weights (LWA) for each attention head. $Pipeline\_Depth\_MHA$ (PD\_MHA) equals ($\frac{d_{model}}{TS_{MHA}}$) plus the time required to load, multiply (2 cc), add (1 cc), and store for \mbox{computing} self-attention (SA) in $QKV_{PM}$ module (Equation~\ref{MHA}). $Pipeline\_Depth\_Bias\_Add$ (PD\_BA) includes latency associated with loading, adding, and storing operations in bias addition (BA) tasks (Equation~\ref{BA_mha}). $Pipeline\_Depth\_Score$ (PD\_S) equals ($\frac{d_{model}}{h}$), the time required to compute the score (S) in $QK_{PM}$ module (Equation~\ref{s}). $Pipeline\_Depth\_SV$ (PD\_SV) equals $Sequence\_Length$ in the computation of SV within the $SV_{PM}$ module (Equation~\ref{sv}). Equation~\ref{soft} estimates time for softmax (SM) calculation, which includes exponentiation (4 cc) and division (14 cc). It starts after the $QK_{PM}$ module is finished.
\begin{multicols}{2}
\noindent\begin{gather}
\label{MHA}
\begin{split}
   SA = [(\frac{d\_{model}}{h}-1)\times 1 + PD\_MHA]\times SL 
\end{split}\\
\label{BA_mha}
\begin{split}
     BA =& [(\frac{d\_{model}}{h}-1)\times 1 + PD\_BA] \times SL 
\end{split}\\
\label{s}
\begin{split}
      Score (S) = [(SL-1)\times 1 + PD\_S]\times SL 
\end{split}\\
\label{sv}
\begin{split}
      SV = [(\frac{d\_{model}}{h}-1)\times 1 + PD\_SV]\times SL 
\end{split}
\end{gather}
\begin{gather}
\label{soft}
\begin{split}
\hfill\\
   SM =& [(SL-1)\times 1 + Load + Store]\times SL + [(SL-\\& 1) \times 1 + Load + Store + add + \\& exponentiation]\times SL+[(SL-1)\times 2+\\&  
    Load + Store + divide]\times SL
\end{split}
\end{gather}
\end{multicols}

\subsection{Latency model for FFN1 Module}
\begin{gather}
\label{LIF}
\begin{split}
  LIF1=[(\frac{d\_{model}}{Tile\ no.\ FFN}-1)\times 1 + PD\_LFFN1]\times SL
\end{split}
\end{gather}

\begin{multicols}{2}
\noindent\begin{gather}
\label{LWF1}
\begin{split}
  LWF1=&[(\frac{d\_{model}}{Tile\ no.\ FFN}-1)\times 1 + PD\_L]\\& \times \frac{d\_{model}}{Tile\ no.\ FFN} 
\end{split}\\
\label{LBF}
\begin{split}
  LBF1=(d\_{model}-1)\times 1 + PD\_L 
\end{split}
\end{gather} 
\begin{gather}
\label{FFN1}
\begin{split}
   FFN1=&[(\frac{d\_{model}}{Tile\ no.\ FFN}-1)\times 1 \\& + PD\_FFN1]\times SL 
\end{split}\\
\label{BAFFN}
\begin{split}
    BAF1=&[(d\_{model}-1)\times 1 + PD\_BA]\\& \times SL 
\end{split}
\end{gather} 
\end{multicols}

where, $Load\_Inputs\_FFN1$ (LIF1) unit loads tiled outputs from the attention module into the input buffer of the FFN1 module. $Load\_Weights\_FFN1$ (LWF1) unit loads partial weights from off-chip memory to the weight buffer of the FFN1 module according to $TS_{FFN}$. $Pipeline\_Depth\_FFN1$ (PD\_FFN1) equals ($\frac{d_{model}}{Tile\ no.\ FFN}$) plus the time required to perform load, add, and store operations in the FFN1 module. $Pipeline\_Depth\_Load\_FFN1$ (PD\_LFFN1) is the time required to load, add, and store in the loading units. FFN1 is the computation time of $FFN1_{PM}$ module. $Load\_Biases\_FFN$ (LBF1) loads biases to registers from off-chip memory while $FFN1_{PM}$ operates. $Bias\_Addition\_FFN1$ (BAF1) adds biases to the outputs of $FFN1_{PM}$.

\subsection{Model for BRAM utilization}
Equation~\ref{eq:bram} gives an estimate for BRAM consumption. It was derived from all the arrays declared in HLS with true dual-port BRAM pragmas. 
\begin{equation}
\label{eq:bram}
\begin{split}
    No.\ of\ BRAMs =& \frac{10\times SL\times d_{model} \times Bit\_w}{BRAM\_w\times BRAM\_d} + SL\ \times max\left(0.5, \frac{SL\times Bit\_w}{BRAM\_w\times BRAM\_d}\right) \\ & +  max\left(0.5, \frac{SL\times d_{model}\times Bit\_w}{BRAM\_w\times BRAM\_d}\right)
            + \frac{h\times SL\times d_{model}\times Bit\_w}{BRAM\_w\times BRAM\_d} \\ & + max\left(0.5, \frac{d_{model}\times Bit\_w}{BRAM\_w\times BRAM\_d}\right)  + \frac{SL\times Tile\ no.\ MHA\times Bit\_w}{BRAM\_w\times BRAM\_d} \\ & +  Tile\ no.\ MHA \times h\times max\left(0.5, \frac{SL\times Bit\_w}{BRAM\_w\times BRAM\_d}\right) \\ & + \frac{8\times d_{model}^2 \times Bit\_w}{Tile\ no.\ FFN\times BRAM\_w\times BRAM\_d} + Tile\ no.\ MHA\times h \\ & \times max\left(0.5, \frac{d_{model}\times Bit\_w}{BRAM\_w\times BRAM\_d}\right) + \frac{d_{model}}{Tile\ no.\ FFN}\times max(0.5, \\ &\left.\frac{SL\times Bit\_w}{BRAM\_w\times BRAM\_d}\right) + 4\times d_{model}\times max\left(0.5, \frac{SL\times Bit\_w}{BRAM\_w\times BRAM\_d}\right)
\end{split} 
\end{equation}

Here, BRAM\_d is the depth of BRAMs, which indicates the number of storage locations (or entries) within a BRAM block. Each location holds a fixed number of bits, defined by the width of BRAM (BRAM\_w), and both parameters can vary depending on the platform. Bit\_w is the bit precision of the data being stored. BRAM\_w = 36 and BRAM\_d = 1024 for most FPGAs. \textcolor{red}{\hl{}}Each term in the equation corresponds to an array declared in the HLS code. For instance, the first term represents the number of BRAMs synthesized for 10 arrays of size SL$\times$ d$_{model}$. The max function in the second term accounts for cases where an array may not fully utilize the 18 kb width of a synthesized BRAM, but at least one 18 kb BRAM will still be allocated. The factor 0.5 is used because the total BRAM count is calculated based on 36 kb BRAMs.

\subsection{Latency model for LN Module}
\begin{multicols}{2}
\noindent\begin{gather}
\label{LWLN}
\begin{split}
   LWN = (d\_{model}-1)\times 1 + PD\_L 
\end{split}
\end{gather}
\begin{gather}
\label{LBLN}
\begin{split}
   LBN = (d\_{model}-1)\times 1 + PD\_L 
\end{split}
\end{gather}
\end{multicols}
\vspace{-1.0cm}
\begin{gather}
\label{RC}
\begin{split}
  RC = [(d\_{model}-1)\times 1 + PD\_BA]\times SL 
\end{split}
\end{gather}
\vspace{-0.8cm}
\begin{equation} 
\label{LN}
\begin{split}
   Layer\ Norm =& [(d\_{model}-1)\times 2 + Load + Add + Store]\times SL+[(d\_{model}-1)\times 2 + Load + multiply \\ & + add + store]\times SL+[(d\_{model}-1)\times 1+Load+Square+multiply+add+Store\\ & +divide+  float\_to\_fixed\_conversion]\times SL+[(d\_{model}-1)\times 1+Load+add\\ & +Store]\times SL
\end{split} 
\end{equation} 

where, $Load\_Weights\_LN$ (LWN) unit loads weights from off-chip memory to the weight buffer of the LN module. $Load\_Biases\_LN$ (LBN) loads biases to registers from off-chip memory. RC represents the operations of the residual connection in LN module. $float\_to\_fixed\_conversion$ in the LN module takes 3 cc.
\subsection{Latency model for FFN2 Module}
\begin{multicols}{2}
\noindent\begin{gather}
\label{LIF2}
\begin{split}
  LIF2=&[(\frac{d_{model}}{Tile\ no.\ FFN}-1)\times 1 \\& + PD\_LFFN2]\times SL 
\end{split}\\
\label{LWF2}
\begin{split}
  LWF2=&[(\frac{d_{model}}{Tile\ no.\ FFN}-1)\times 1 + PD\_L]\\& \times \frac{d_{model}}{Tile\ no.\ FFN} 
\end{split}
\end{gather} 
\begin{gather}
\label{LBF2}
\begin{split}
  LBF2=(d_{model}-1)\times 1+PD\_L 
\end{split}\\
\label{FFN2}
\begin{split}
   FFN2=&[(\frac{4\times d_{model}}{Tile\ no.\ FFN}-1)\times 1\\& +PD\_FFN2]\times SL 
\end{split}\\
\label{BAFFN2}
\begin{split}
    BAF2=&[(4\times d_{model}-1)\times 1+PD\_BA]\\& \times SL 
\end{split}
\end{gather} 
\end{multicols}
where, $Load\_Inputs\_FFN2$ (LIF2) unit loads tiled outputs from the FFN1 module into the input buffer of the FFN2 module. $Load\_Weights\_FFN2$ (LWF2) unit loads partial weights from off-chip memory to the weight buffer of the FFN2 module according to $TS_{FFN}$. $Pipeline\_Depth\_FFN2$ equals ($\frac{d_{model}}{Tile\ no.\ FFN}$) plus the time required to perform load, add, and store operations in the FFN2 module. $Pipeline\_Depth\_Load\_FFN2$ (PD\_LFFN2) is the time required to load, add, and store in the loading units. FFN2 is the computation time of $FFN2_{PM}$ module. $Load\_Biases\_FFN2$ (LBF2) loads biases to registers from off-chip memory while $FFN2_{PM}$ operates. $Bias\_Addition\_FFN2$ (BAF2) adds biases to the outputs of $FFN2_{PM}$.

\subsection{Latency model for FFN3 Module}
\begin{multicols}{2}
\noindent\begin{gather}
\label{LIF3}
\begin{split}
  LIF3=&[(\frac{4\times d_{model}}{Tile\ no.\ FFN}-1)\times 1 \\& + PD\_LFFN3]\times SL 
\end{split}\\
\label{LWF3}
\begin{split}
  LWF3=&[(\frac{4\times d_{model}}{Tile\ no.\ FFN}-1)\times 1 + PD\_L]\\& \times \frac{d_{model}}{Tile\ no.\ FFN} 
\end{split}
\end{gather} 
\begin{gather}
\label{LBF3}
\begin{split}
  LBF3=(d_{model}-1)\times 1 + PD\_L 
\end{split}\\
\label{FFN3}
\begin{split}
   FFN3=&[(\frac{d_{model}}{Tile\ no.\ FFN}-1)\times 1 \\&+ PD\_FFN3]\times SL 
\end{split}\\
\label{BAFFN3}
\begin{split}
    BAF3=&[(d_{model}-1)\times 1 + PD\_BA]\\& \times SL  
\end{split}
\end{gather} 
\end{multicols}
where, $Load\_Inputs\_FFN3$ (LIF3) unit loads tiled outputs from the FFN2 module into the input buffer of the FFN3 module. $Load\_Weights\_FFN3$ (LWF3) unit loads partial weights from off-chip memory to the weight buffer of the FFN3 module according to $TS_{FFN}$. $Pipeline\_Depth\_FFN3$ equals ($\frac{4\times d_{model}}{Tile\ no.\ FFN}$) plus the time required to perform load, add, and store operations in the FFN3 module. $Pipeline\_Depth\_Load\_FFN3$ is the time required to load, add, and store in the loading units. FFN3 is the computation time of $FFN3_{PM}$ module. $Load\_Biases\_FFN3$ (LBF3) loads biases to registers from off-chip memory while $FFN3_{PM}$ operates. $Bias\_Addition\_FFN3$ (BAF3) adds biases to the outputs of $FFN3_{PM}$.

\section{Evaluation and Results}\label{results}


\textbf{\textit{ADAPTOR}} supports software-level programmability, allowing modification of key design parameters at runtime. These parameters include the embedding dimension ($d_{model}$), number of attention heads (h), number of encoder layers (N), and sequence length (SL). Initially, these parameters were configured with fixed values of 768, 12, 12, and 64, respectively, based on a BERT variant \cite{bert}, which is a widely used transformer model for natural language processing, and the available FPGA resources. In contrast, the tile sizes are fixed at synthesis and cannot be modified at runtime. Consequently, synthesis was performed with fixed tile sizes of $TS_{MHA} = 64$ and $TS_{FFN} = 128$. This design approach is a key contribution that gives \textbf{\textit{ADAPTOR}} 
the ability to retain a single, resource-constrained synthesis configuration while enabling runtime configurability of core transformer parameters so that it can support diverse transformer neural network models without requiring re-synthesis.


\begin{figure}
\centering
    \begin{subfigure}{.5\textwidth}
    \centering
    \includegraphics[height=5cm, width=1.0\linewidth]{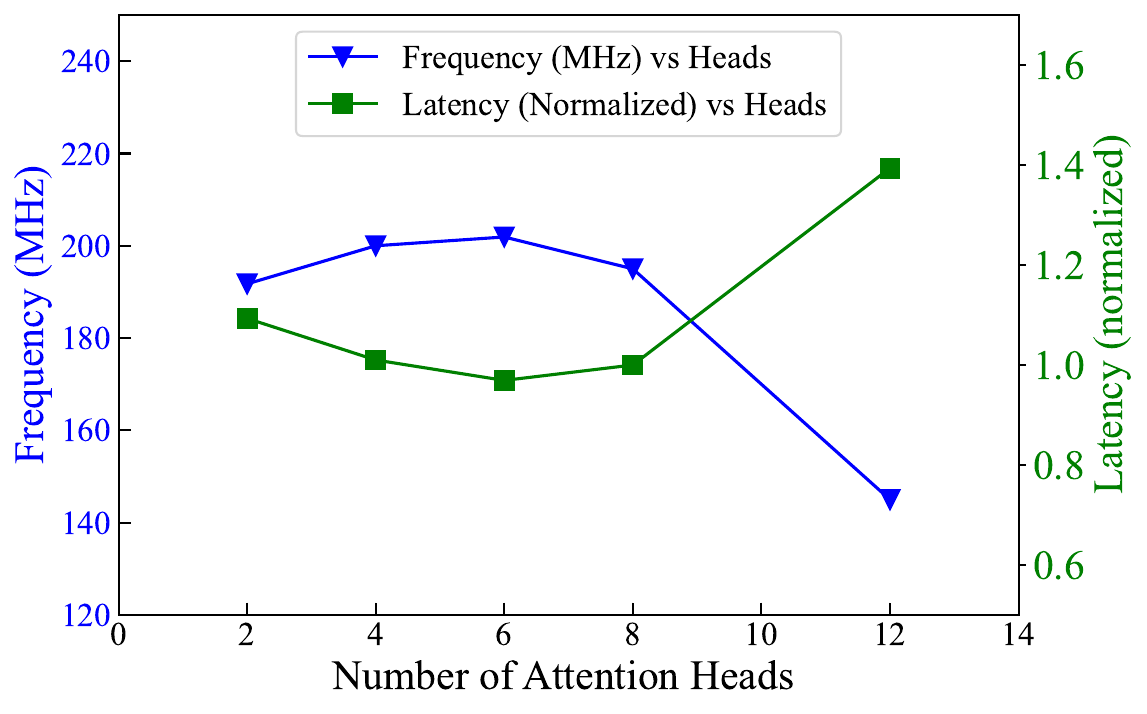}
    \caption{\label{head_perf} Variation of Performance with number of attention heads.}
    \end{subfigure}%
    \begin{subfigure}{.5\textwidth}
    \centering
    \captionsetup{justification=centering}
    \includegraphics[height=5cm, width=1.0\linewidth]{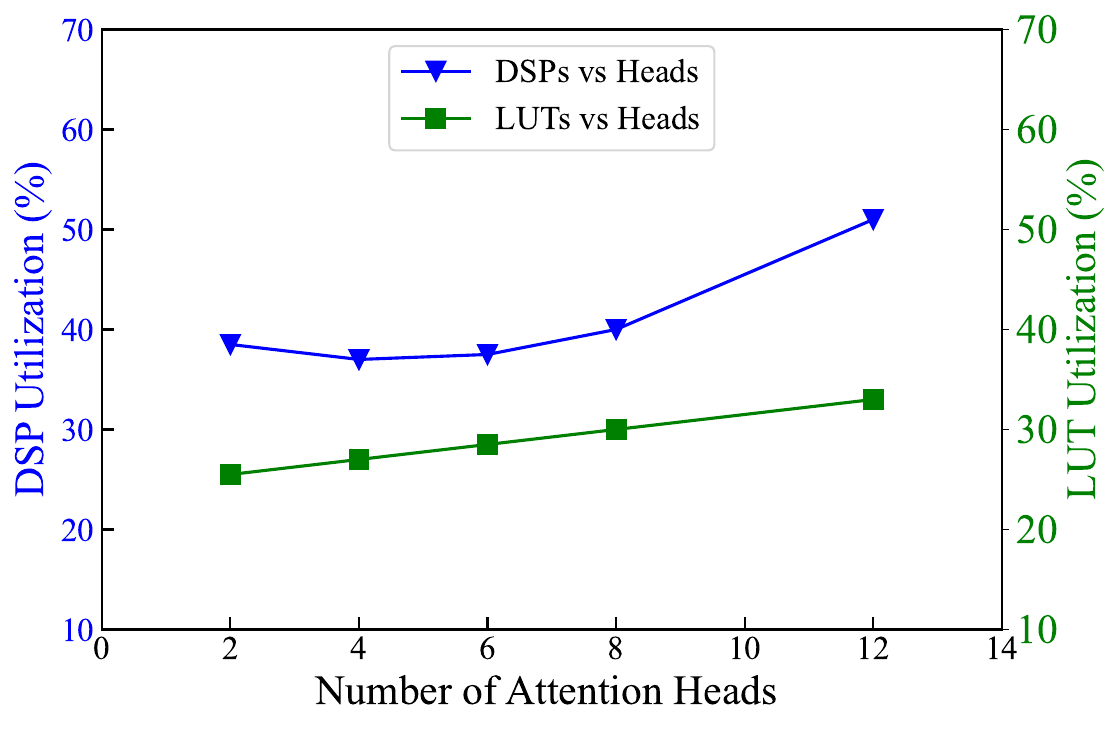}
    \caption{\label{head_util}Variation of Resources with number of attention heads.}
    \end{subfigure}
\caption{Performance and resource utilization VS attention heads.}
\end{figure}


Fig.~\ref{head_perf} presents the effect of varying the number of attention heads on system frequency and normalized latency, where latency accounts for computation time assuming overlap with data loading. While increasing the number of attention heads generally improves parallelism and reduces latency, the system frequency decreases beyond a certain threshold, leading to higher latency. Optimal performance is observed with 6–10 attention heads. Fig.~\ref{head_util} illustrates the corresponding increase in DSP and LUT utilization, showing that higher resource usage contributes to reduced system frequency. These results provide a quantitative analysis of the trade-off between parallelism and hardware timing constraints, identify an optimal design point for FPGA-based transformers, and characterize the impact of resource utilization on latency and frequency. Furthermore, the evaluation methodology accounts for overlapped data loading and computation, offering a realistic performance assessment for hardware accelerators.

\begin{figure}
\centering
\captionsetup{justification=centering}
    \includegraphics[height=7cm, width=1.0\linewidth]{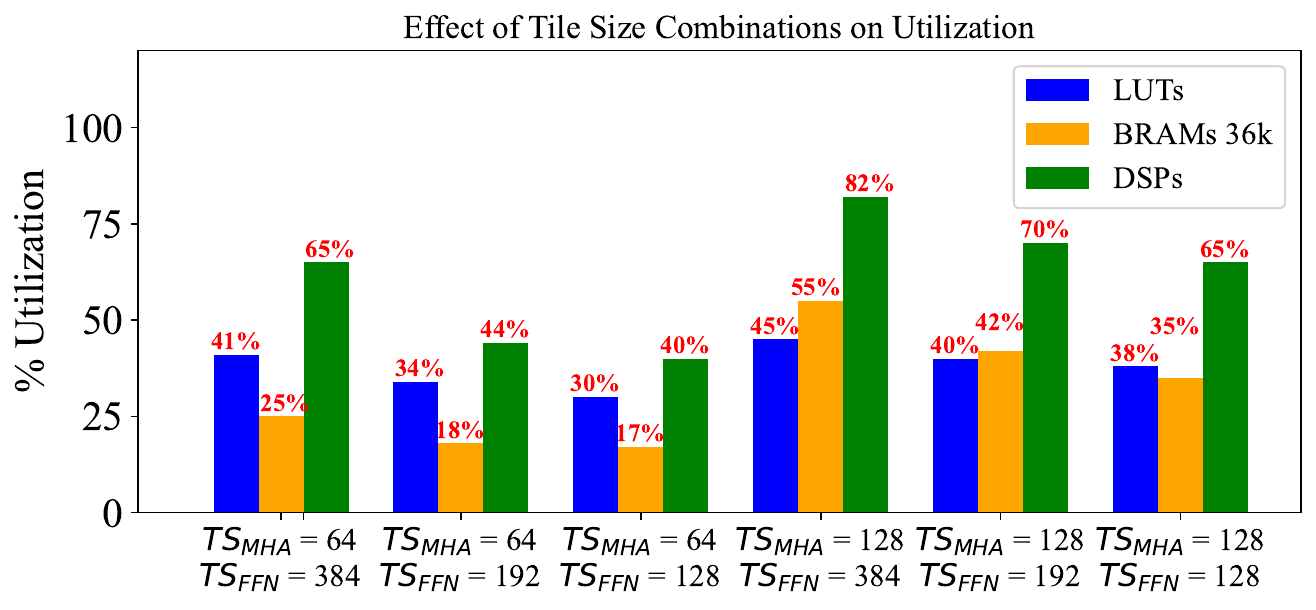}
    \caption{\label{util_tile} Utilization vs tile size.}
\end{figure}


Fig.~\ref{util_tile} illustrates the effect of varying tile sizes ($TS_{MHA}$, $TS_{FFN}$) on the utilization of DSP, LUT, and BRAM. Since processing modules rely on DSPs for multiplication–accumulation (MAC) operations, DSPs represent the most widely used resource and can reach saturation before BRAMs, rendering accelerator computation bound. Increasing the tile sizes for both the attention and feedforward modules results in higher DSP utilization, which enables greater parallelism and reduces latency until the system frequency begins to decline. This analysis is the characterization of resource–performance trade-offs, showing that tile size selection directly determines when the accelerator becomes computation-bound or frequency-limited, thus guiding optimal design choices for FPGA-based transformer implementations.

\begin{figure}
\centering
\captionsetup{justification=centering}
    \includegraphics[height=8cm, width=1.0\linewidth]{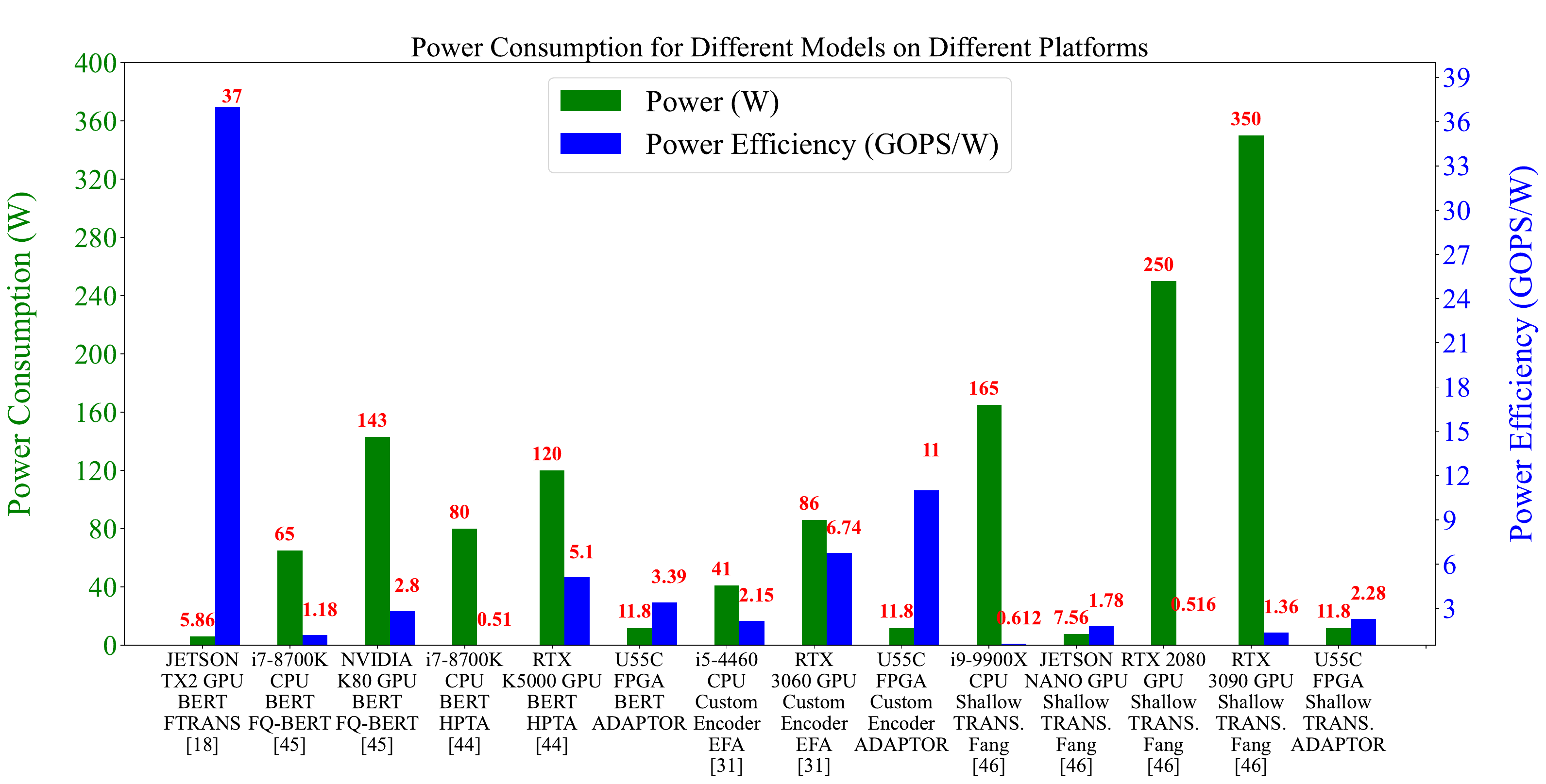}
    \caption{\label{power} Cross Platform Comparison of Power Consumption.}
\end{figure}

Fig.~\ref{power} compares the power consumption (in watts) and power efficiency (throughput per watt, GOPS/W) for various models across different CPUs, GPUs, and our FPGA accelerator. Data for different models and platforms were obtained from cited literature, and we used them to compare the performance of \textbf{\textit{ADAPTOR}} on the U55C platform for the same models. Since \textbf{\textit{ADAPTOR}} is synthesized only once, and power is measured using Vivado's power estimation tool post-synthesis, the total dynamic power consumption remains constant for all models. The JETSON TX2 GPU \cite{li_ftrans_2020} achieves the highest power efficiency for the BERT model, mainly due to the sparse architecture of the algorithm, and also has the lowest overall power consumption. The RTX K5000 GPU \cite{han_hpta_nodate} is 1.5$\times$ more power efficient than \textbf{\textit{ADAPTOR}} for the BERT model, due to compression techniques, but consumes 10$\times$ more power. The i7-8700K CPU is the least power-efficient for BERT \cite{han_hpta_nodate}. \textbf{\textit{ADAPTOR}} is 1.2$\times$ and 2.87$\times$ more power efficient than the NVIDIA K80 GPU and i7-8700K CPU, respectively, when running BERT, according to FQ-BERT \cite{liu_hardware_2021}. A custom encoder with four encoding layers was run on an i5-4460 CPU and an RTX 3060 GPU \cite{yang_efa-trans_2022}, both of which were 5.1$\times$ and 1.63$\times$ less power efficient than \textbf{\textit{ADAPTOR}} while also being more power-hungry. Fang et al.\cite{fang_algorithm} executed a shallow transformer on an i9-9900X CPU, JETSON NANO GPU, RTX 2080, and RTX 3090 GPUs. Although the JETSON NANO GPU consumed 1.56$\times$ less power than \textbf{\textit{ADAPTOR}}, the other devices used 14–30$\times$ more power. However, \textbf{\textit{ADAPTOR}} is 3.7$\times$, 1.28$\times$, 4.4$\times$, and 1.67$\times$ more power efficient than all of them.

\begin{figure}
\centering
    \centering
    \captionsetup{justification=centering}
    \includegraphics[height=8.0cm, width=0.7\linewidth]{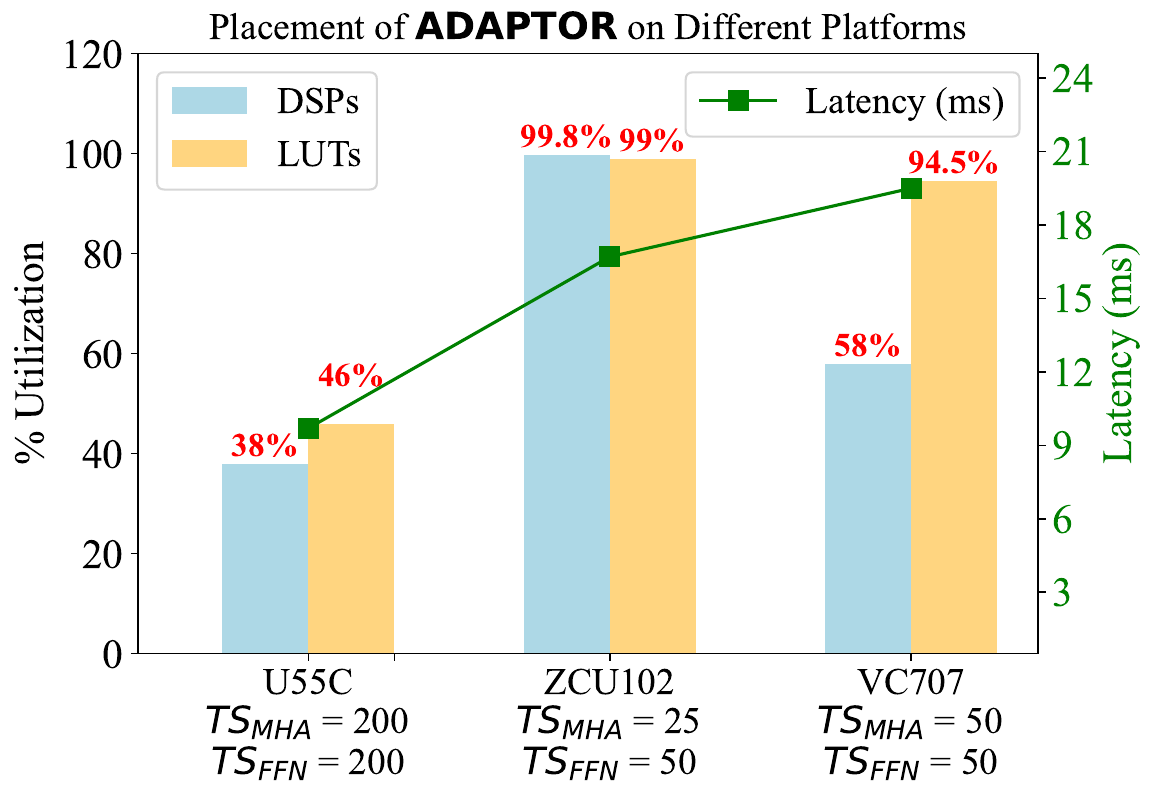}
    \caption{\label{port} Testing Portability Feature.}
\end{figure}

Fig.~\ref{port} illustrates that \textbf{\textit{ADAPTOR}} can be deployed on any platform, regardless of the size of the TNN model or available resources, by adjusting the $TS_{MHA}$ and $TS_{FFN}$ parameters in HLS during design time. The figure presents results for a custom TNN encoder with an embedding dimension of 200, 3 attention heads, 2 encoder layers, and a sequence length of 64. On the Alveo U55C, the tile sizes can be maximized ($TS_{MHA}$ = 200, $TS_{FFN}$ = 200) due to the abundance of resources, resulting in lower latency. For the ZCU102 board, the tile sizes were reduced to 25 and 50 respectively, to fit the model within its resource constraints, nearly consuming 100\% of the DSPs and LUTs and increasing the latency. On the VC707 board, $TS_{MHA}$ and $TS_{FFN}$ were set to 50 each, as it has slightly more resources than the ZCU102. However, latency increased as fewer DSPs were utilized, and LUT consumption almost reached its limit. 

\begin{figure}
\centering
    \centering
    \captionsetup{justification=centering}
    \includegraphics[height=8cm, width=0.8\linewidth]{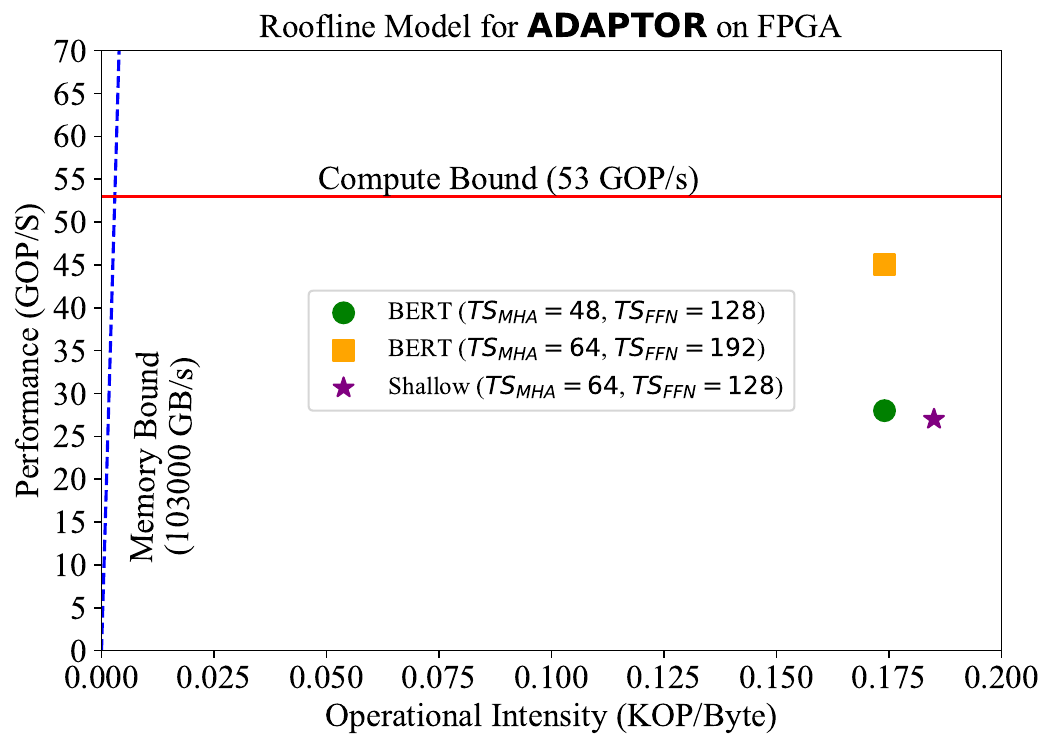}
    \caption{\label{roof}Peak Performance and Peak Memory Bandwidth.}
\end{figure}

Fig.~\ref{roof} \textcolor{red}{\hl{}}presents the roofline model of \textbf{\textit{ADAPTOR}}, highlighting its peak performance and memory bandwidth limits. The \textit{Memory Bound} (blue dashed line) indicates the maximum achievable performance based on the memory bandwidth, which is 103,000 GB/s. Data points to the left of this line are constrained by memory bandwidth. The \textit{Compute Bound} (red line) represents the peak performance determined by the FPGA's computational resources, capped at 53 GOP/s. Points below this line indicate underutilization of computational resources. All data points (green, yellow, and purple) fall within the compute and memory-bound regions, meaning none fully utilize the accelerator's available resources. The yellow square, representing the BERT model with $TS_{MHA} = 64$ and $TS_{FFN} = 192$, achieves the highest performance, being closest to the compute bound. In contrast, the purple star, corresponding to the shallow transformer model with $TS_{MHA} = 64$ and $TS_{FFN} = 128$, exhibits the highest operational intensity but the lowest performance.

Equation \ref{bw} below is used to calculate memory bandwidth (BW), where no.\ of\ BRAMs $=$ 340, BRAM's\ width $=$ 36 KB, no.\ of\ LUTRAMs $=$ 129101, LUTRAM's\ Width $=$ 32 KB. Our previous work \cite{protea} calculated latency and throughput data (53 GOP/s).

\noindent\begin{gather}
\label{bw}
\begin{split}
    Memory\ Bandwidth = & (No.\ of\ BRAMs \times BRAM's\ Width + No.\ of\ LUTRAMs \\& \times LUTRAM's\ Width) \times Frequency
\end{split}
\end{gather}

\begin{table}[htbp]
\setlength{\arrayrulewidth}{0.05pt}
\renewcommand{\arraystretch}{1}
  \centering
  \caption{Comparison with FPGA Accelerators.}
    \begin{tabular}{|c|c|c|c|c|c|c|c|c|c|}
    \hline
    \multicolumn{1}{|c|}{\multirow{2}[2]{*}{\textbf{Accelerator}}} & \multirow{2}[2]{*}{\textbf{DSP}} & \multirow{2}[2]{*}{\textbf{LUT}}  & \multirow{2}[2]{*}{\textbf{GOPS}} & \textbf{Power} & \textbf{(GOPS/DSP)} & \textbf{(GOPS/LUT)} & \textbf{GOPS/} & \multirow{2}[2]{*}{\textbf{Method}} & \multirow{2}[2]{*}{\textbf{Sparsity}} \bigstrut[t]\\
    \multicolumn{1}{|c|}{} & \multicolumn{1}{c|}{} &   &   & \textbf{(W)}  & \textbf{$\times$1000}  &  \textbf{$\times$1000}  & \textbf{Power}  &  &  \bigstrut[b]\\
    \hline 
    \multicolumn{1}{|c|}{Network \#1} & \multicolumn{9}{c|}{Shallow Transformer}   \bigstrut\\
    \hline
    \multicolumn{1}{|c|}{Qi et al. \cite{qi_accommodating_2021}} & 3572 (52\%) & 485 k (41\%) & 14 & -- & 3.92 & 0.03 & -- & \multirow{3}[3]{*}{HLS} & {80\%} \bigstrut\\ 
\cline{1-8} \cline{10-10}
    \multicolumn{1}{|c|}{Qi et al. \cite{qi_accelerating_2021}} & 5040 (74\%)  & 908 k (76\%) & 12 & -- & 2.38 & 0.013 & -- &  & {86\%} \bigstrut\\ 
\cline{1-8} \cline{10-10}  \textbf{\textit{ADAPTOR}} & 3612 (40\%) & 391 k (30\%) & 27 & 11.8 & 7.47 & 0.069 & 2.28 &  & 0\% \bigstrut\\
    \hline
    \multicolumn{9}{c}{} \bigstrut\\ [-1.3em]
    \hline
    \multicolumn{1}{|c|}{Network \#2} & \multicolumn{9}{c|}{Custom Transformer Encoder}   \bigstrut\\
    \hline
    \multicolumn{1}{|c|}{Qi et al. \cite{qi_accelerating_2021}} & 4145 (60\%) & 937 k (79\%) & 75.94 & -- & 18 & 0.08 & -- & \multirow{2}[2]{*}{HLS} & \multirow{2}[2]{*}{0\%} \bigstrut\\ 
\cline{1-8}   \textbf{\textit{ADAPTOR}} & 3612 (40\%) & 391 k (30\%) & 132 & 11.8 & 37 & 0.34 & 11  &  &  \bigstrut\\
    \hline
    \multicolumn{9}{c}{} \bigstrut\\ [-1.3em]
    \hline
    \multicolumn{1}{|c|}{Network \#3} & \multicolumn{9}{c|}{BERT}   \bigstrut\\
    \hline
    \multicolumn{1}{|c|}{Tzanos et al. \cite{tzanos_hardware_2022}} & 5861 (85\%) & 910 k (77\%) & 65.7 & -- & 11.2 & 0.07 & -- & {} & 0\% \bigstrut\\ 
\cline{1-8} \cline{10-10}
    \multicolumn{1}{|c|}{TRAC \cite{plagwitz_trac_2022}} & 1379 (80\%) & 126 k (55\%) & 128 & -- & 93 & 1.01 & -- & {} & -- \bigstrut\\ 
\cline{1-8} \cline{10-10}   \textbf{\textit{ADAPTOR}} & 3612 (40\%) & 391 k (30\%) & 40 & 11.8 & 11 & 0.10 & 3.39  & {} & 0\% \bigstrut\\

\hline
    \end{tabular}%
  \label{fpga}%
\end{table}%

Table~\ref{fpga} compares the performance of our accelerator, \textbf{\textit{ADAPTOR}}, with other FPGA-based accelerators. Each of these accelerators is optimized for specific TNN models, with some designed for sparse computations. TRAC \cite{plagwitz_trac_2022} is the only one that automatically generates accelerator code based on the target FPGA and TNN architecture. Since \textbf{\textit{ADAPTOR}} was synthesized once with fixed hardware resources and bit width, and implemented on a dense model without sparsity, we evaluated throughput (GOPS), power consumption, normalized throughput (GOPS per DSP or GOPS per LUT), and power efficiency (GOPS per watt) for a fair comparison. \textbf{\textit{ADAPTOR}} achieved 1.9$\times$ and 2.25$\times$ higher GOPS compared to the accelerators by Qi et al. in \cite{qi_accommodating_2021} and \cite{qi_accelerating_2021}, respectively, for a shallow transformer. Its normalized throughput was also higher, indicating more efficient DSP and LUT usage without relying on pruning, whereas Qi et al. employed block balanced pruning and block row storage. Qi et al.'s four-layer transformer encoder \cite{qi_accelerating_2021} was 1.7$\times$ slower and 2$\times$ less resource-efficient than \textbf{\textit{ADAPTOR}} even with hierarchical pruning. TRAC \cite{plagwitz_trac_2022} consumed fewer DSPs and LUTs but reported 3.2$\times$ higher GOPS and 8.4$\times$ higher GOPS/DSP. None of these accelerators incorporated tiling or partitioning schemes to support large models such as BERT, which our design explicitly addresses. Tzanos et al. \cite{tzanos_hardware_2022} applied tiling and used more resources, achieving 1.6$\times$ higher speed with GOPS/DSP comparable to \textbf{\textit{ADAPTOR}}.

\begin{figure}
\centering
    \centering
    \captionsetup{justification=centering}
    \includegraphics[height=7.0cm, width=1.0\linewidth]{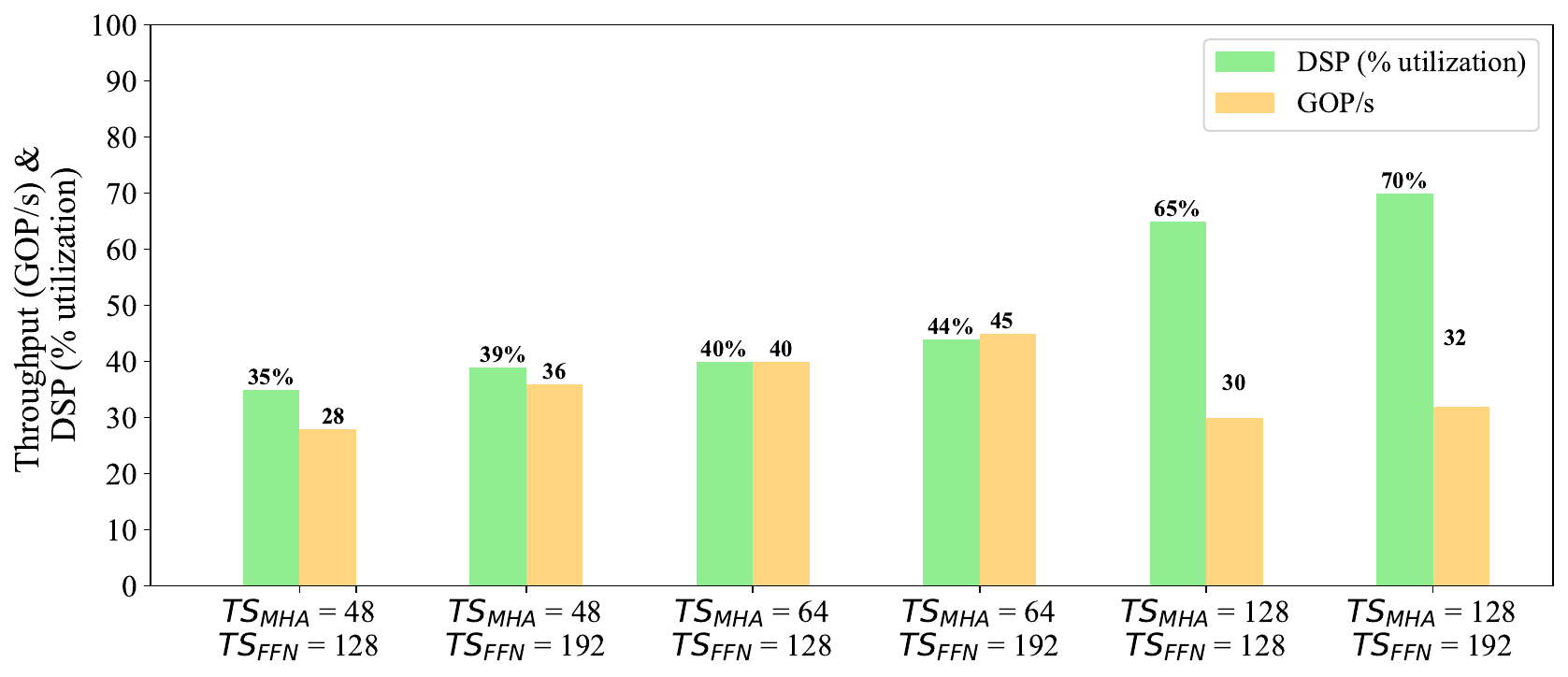}
    \caption{\label{GOPS_DSP} Effect of DSP on GOP/s for Various Tile Size Combinations.}
\end{figure}

Although \textbf{\textit{ADAPTOR}} utilizes over 3,000 DSPs at 200 MHz, the measured throughput is significantly lower than the theoretical peak of ~1200 GOPS. This underutilization arises from several factors: (i) sequential execution of certain submodules (e.g., $QKV_{PM}$, $QK_{PM}$, $SV_{PM}$, $FFN1_{PM}$, $FFN2_{PM}$, $FFN3_{PM}$, softmax, layer normalization etc.) that prevents simultaneous activation of all functional modules, and (ii) control dependencies that limit pipelining across nested loops. As a result, a portion of the available DSPs remain idle at different stages of execution, reducing overall efficiency. Fig.~\ref{GOPS_DSP} illustrates how GOPS scales with DSP consumption as the tile sizes of the MHA and FFN layers increase. While larger tiles increase DSP utilization and improve throughput, the system frequency drops beyond certain tile sizes (Fig.~\ref{tile_select}), leading to diminishing returns and even a reduction of GOPS to 30 and 32 for 65\% and 70\% DSP utilization, respectively. This analysis highlights the fundamental trade-off between resource utilization, frequency, and achievable throughput in FPGA-based transformer accelerators.

\begin{table*}[htbp]
\setlength{\arrayrulewidth}{0.05pt}
\renewcommand{\arraystretch}{1.0}
  \centering
  \caption{Validation of Experimental and Analytical Results}
    \begin{tabular}{|c|c|c|c|c|c|c|c|c|c|c|c|}
    \hline
    \multirow{4}[4]{*}{\textbf{Method}} & \multicolumn{1}{c|}{\multirow{3}[2]{*}{\textbf{Sequence}}} & \multicolumn{1}{c|}{\multirow{3}[2]{*}{\textbf{Embedding}}} & \multicolumn{1}{c|}{\multirow{2}[2]{*}{\textbf{Number}}} & \multicolumn{1}{c|}{\multirow{2}[1]{*}{\textbf{Tile}}} & \multicolumn{1}{c|}{\multirow{2}[1]{*}{\textbf{Tile}}} & \multicolumn{1}{c|}{\multirow{4}[4]{*}{\textbf{DSPs}}} & \multicolumn{1}{c|}{\multirow{3}[3]{*}{\textbf{BRAMs}}} & \multicolumn{1}{c|}{\multirow{3}[3]{*}{\textbf{Frequency}}} & \multicolumn{3}{c|}{\textbf{Latency (ms)}} \bigstrut\\
\cline{10-12}    \multicolumn{1}{|c|}{} & \multicolumn{1}{c|}{\multirow{3}[2]{*}{\textbf{Length}}} & \multicolumn{1}{c|}{\multirow{3}[2]{*}{\textbf{Dimension}}} & \multicolumn{1}{c|}{\multirow{2}[-1]{*}{\textbf{of}}} & \multicolumn{1}{c|}{\multirow{2}[-1]{*}{\textbf{Size}}} & \multicolumn{1}{c|}{\multirow{2}[-1]{*}{\textbf{Size}}} & \multicolumn{1}{c|}{} & \multirow{3}[3]{*}{\textbf{18k}} &  \multirow{3}[3]{*}{\textbf{(MHz)}}  & \multicolumn{1}{c|}{\textbf{Attention}} & \multicolumn{1}{c|}{\textbf{Load}} & \textbf{FFN} \bigstrut[t]\\
    \multicolumn{1}{|c|}{} & \multicolumn{1}{c|}{} & \multicolumn{1}{c|}{} & \multicolumn{1}{c|}{\multirow{2}[2]{*}{\textbf{Heads}}} & \multicolumn{1}{c|}{\multirow{2}[2]{*}{\textbf{MHA}}} & \multicolumn{1}{c|}{\multirow{2}[2]{*}{\textbf{FFN}}} & \multicolumn{1}{c|}{} &   &   & \multicolumn{1}{c|}{\textbf{Module}} & \multicolumn{1}{c|}{\textbf{Weights Unit}} &  \textbf{Module} \\
    \multicolumn{1}{|c|}{} & \multicolumn{1}{c|}{} & \multicolumn{1}{c|}{} & \multicolumn{1}{c|}{} & \multicolumn{1}{c|}{} & \multicolumn{1}{c|}{} & \multicolumn{1}{c|}{} &   &   & \multicolumn{1}{c|}{\textbf{(SA)}} & \multicolumn{1}{c|}{\textbf{(LWA)}} &  \textbf{(FFN1)} \bigstrut[b]\\
    \hline
    Analytical & \multirow{2}[2]{*}{64} & \multirow{2}[2]{*}{768} & \multirow{2}[2]{*}{8} & \multirow{2}[2]{*}{64} & \multirow{2}[2]{*}{128} & 3784 & 2375 & \multirow{6}[10]{*}{200} & 0.052 & 0.037 & 0.082 \bigstrut[t]\\
\cline{1-1} Experimental &   &   &   &   &   & 3612 & 2246 &   & 0.053 & 0.038 & 0.084 \bigstrut[b]\\
\cline{1-8}\cline{10-12}    \multicolumn{8}{|c|}{}        &   & \multicolumn{3}{c|}{} \bigstrut\\ [-1.5em]
\cline{1-8}\cline{10-12}    Analytical & \multirow{2}[2]{*}{128} & \multirow{2}[2]{*}{768} & \multirow{2}[2]{*}{8} & \multirow{2}[2]{*}{64} & \multirow{2}[2]{*}{128} & 3784 & 2375 &   & 0.103 & 0.037 & 0.165 \bigstrut[t]\\
\cline{1-1} Experimental &   &   &   &   &   & 3612 & 2246 &   & 0.106 & 0.038 & 0.168 \bigstrut[b]\\
\cline{1-8}\cline{10-12}    \multicolumn{8}{|c|}{}        &   & \multicolumn{3}{c|}{} \bigstrut\\ [-1.5em]
\cline{1-8}\cline{10-12}    Analytical & \multirow{2}[2]{*}{64} & \multirow{2}[2]{*}{512} & \multirow{2}[2]{*}{8} & \multirow{2}[2]{*}{64} & \multirow{2}[2]{*}{128} & 3784 & 2375 &   & 0.042 & 0.025 & 0.055 \bigstrut[t]\\
\cline{1-1} Experimental &   &   &   &   &   & 3612 & 2246 &   & 0.043 & 0.026 & 0.056 \bigstrut[b]\\ 
    \hline
    \multicolumn{12}{|c}{} \bigstrut\\ [-1.5em]
    \hline
    Analytical & \multirow{2}[0]{*}{64} & \multirow{2}[0]{*}{768} & \multirow{2}[0]{*}{8} & \multirow{2}[0]{*}{128} & \multirow{2}[0]{*}{192} & 6272 & 2955 & \multirow{2}[0]{*}{135} & 0.11 & 0.1 & 0.18 \\
\cline{1-1} Experimental &   &   &   &   &   & 6317 & 1693 &   & 0.11 & 0.1 & 0.23 \\ 
    \hline
    \end{tabular}%
  \label{validate}%
\end{table*}%


Table~\ref{validate} presents a comparison between the experimental results of \textbf{\textit{ADAPTOR}} and the theoretical predictions derived in Section~\ref{analyt}. For clarity, only a subset of design configurations is reported, focusing on the computation time of the attention and feedforward modules as well as the loading time of the attention module. Latency is primarily influenced by parameters such as sequence length, embedding dimension, and number of attention heads. The measured latency closely aligned with the theoretical estimates, with an average deviation of only 1.8\%. Resource utilization remained stable across configurations with fixed tile sizes, whereas variations in tile size led to corresponding changes in both analytical and experimental values. The deviations were relatively small for DSPs (0.71–4.7\%), but larger for BRAMs (5.7–74\%), particularly at larger tile sizes. The latter discrepancy arises because LUTRAMs were increasingly used in place of BRAMs to sustain higher operating frequency, thereby reducing the accuracy of BRAM utilization estimates.
\nocite{yang_efa-trans_2022}
\nocite{han_hpta_nodate}

\section{Conclusion}\label{conclude} 
In this article, we present a runtime-adaptive FPGA-based accelerator for the encoder and decoder layers of transformer neural networks (TNN), designed using a high-level synthesis (HLS) tool. The architecture leverages FPGA parallelism as well as the inherent parallel nature of TNNs. We demonstrated its deployment on various FPGA platforms, including Alveo U55C, VC707, and ZCU102, highlighting how resources like DSPs and LUTs can be effectively utilized to maximize parallelism and minimize latency in HLS designs. The accelerator is software-programmable, enabling adaptability to different topologies without requiring new code generation or re-synthesis. We implemented an efficient tiling technique and data-loading method for weight matrices, ensuring portability and resource-efficient execution across different TNN models. Experimental results indicate that our design outperforms certain CPUs and GPUs in terms of dynamic power consumption and power efficiency, despite no algorithmic optimizations. Moreover, it achieved a 1.7 to 2.25× speedup over leading FPGA-based accelerators. An analytical model was also developed to validate the experimental findings.

\section{Acknowledgments}
This material is based upon work supported by the National Science Foundation under Grant No. 1956071.

\section{Supplementary Materials}\label{sm}
\hspace{-0.4cm}\begin{minipage}{0.45\textwidth}
\begin{algorithm}[H]
\centering
\caption{Load Weights for MHA}\label{loadwq}
\begin{algorithmic}[1]
\For{$(i=1;i<=\frac{Embedding \ Dimension}{Number \ of \ Heads};i=i+1)$}
\State \textbf{\#pragma HLS pipeline off}
\For{$(j=1;j<=Tiles\_in\_MHA;j=j+1)$}
\State \textbf{\#pragma HLS pipeline II = 1}
    \State $W_Q[i][j] \gets weights\_Q[index];$
    \State $W_K[i][j] \gets weights\_K[index];$
     \State $W_V[i][j] \gets weights\_V[index];$
    \State $index \gets index+1;$
\EndFor
\EndFor
\end{algorithmic}
\end{algorithm}
\end{minipage}
\hfill \ \ \ \ \
\begin{minipage}{0.45\textwidth}
\begin{algorithm}[H]
\centering
\caption{Load Inputs for MHA}\label{loadin_MHA}
\begin{algorithmic}[1]
\For{$(i = 1; i <= sequence\_length; i = i + 1)$}
\State \textbf{\#pragma HLS pipeline off}
\For{$(j=1;j<=Tiles\_in\_MHA;j=j+1)$}
\State \textbf{\#pragma HLS pipeline II = 1}
    \State $X_1[i][j] \gets input\_token[index];$
    \State $X_2[i][j] \gets input\_token[index];$
    \State .....................................;
    \State $X_N[i][j] \gets input\_token[index];$ 
    \State $index \gets index+1;$
\EndFor
\EndFor
\end{algorithmic}
\end{algorithm}
\end{minipage}
\hspace{-0.4cm}\begin{minipage}{0.45\textwidth}
\begin{algorithm}[H]
\centering
\caption{Load Inputs for FFN1}\label{loadin_FFN1}
\begin{algorithmic}[1]
\For{$(i=1;i<=sequence\_length;i=i+1$}
\State \textbf{\#pragma HLS pipeline off}
\For{$(j=1;j<TS_{FFN};j=j+1)$)}
\State \textbf{\#pragma HLS pipeline II = 1}
    \State $k \gets (index)*(factor);$
    \State $X_1[i][j] \gets outputs\_MHA[i][k+j];$ 
    \State $X_2[i][j] \gets outputs\_MHA[i][k+j];$
    \State .....................................;
    \State $X_N[i][j] \gets outputs\_MHA[i][k+j];$ 
    \State $index \gets index+1;$
\EndFor
\EndFor
\end{algorithmic}
\end{algorithm}
\end{minipage}
\hfill \ \ \ \ \
\begin{minipage}{0.45\textwidth}
\begin{algorithm}[H]
\centering
\caption{Load Inputs for FFN2 \& FFN3}\label{loadin_FFN3}
\begin{algorithmic}[1]
\For{$(i=1;i<=sequence\_length;i=i+1$}
\State \textbf{\#pragma HLS pipeline off}
\For{$(j=1;j<TS_{FFN};j=j+1)$)}
\State \textbf{\#pragma HLS pipeline II = 1}
    \State $k \gets (index)*(factor);$
    \State $X_1[i][j] \gets outputs\_FFN2[i][k+j];$ 
    \State $X_2[i][j] \gets outputs\_FFN2[i][k+j];$
    \State .....................................;
    \State $X_N[i][j] \gets outputs\_FFN2[i][k+j];$ 
    \State $index \gets index+1;$
\EndFor
\EndFor
\end{algorithmic}
\end{algorithm}
\end{minipage}

\hspace{-0.4cm}\begin{minipage}{0.45\textwidth}
\begin{algorithm}[H]
\centering
\caption{Load Biases for MHA}\label{loadbiasMHA}
\begin{algorithmic}[1]
\For{$(i=1;i<=\frac{Embedding \ Dimension}{Number \ of \ Heads};i=i+1)$}
\State \textbf{\#pragma HLS pipeline II = 1}
    \State $b_q[i] \gets bias\_Q[index];$
    \State $b_k[i] \gets bias\_K[index];$
    \State $b_v[i] \gets bias\_V[index];$
    \State $index \gets index+1;$
\EndFor
\end{algorithmic}
\end{algorithm}
\end{minipage}
\hfill \ \ \ \
\begin{minipage}{0.45\textwidth}
\begin{algorithm}[H]
\centering
\caption{Load Biases for FFN \& Layer Norm.}\label{loadbiasAll}
\begin{algorithmic}[1]
\For{$(i=1;i<=Embedding \ Dimension;i=i+1)$}
\State \textbf{\#pragma HLS pipeline II = 1}
    \State $b_{FFN}[i] \gets bias\_port[index];$
    \State $index \gets index+1;$
\EndFor
\end{algorithmic}
\end{algorithm}
\end{minipage}

\begin{algorithm}[H]
\caption{Softmax}\label{soft_alg}
\begin{minipage}[b]{0.4\linewidth}
\begin{algorithmic}
\Statex{\textbf{Max Value:}}
\For{$(i=1;i<=SL;i++)$}
\State \textbf{\#pragma HLS pipeline off}
\For{$(j=1;j<=SL;j++)$}
\State \textbf{\#pragma HLS pipeline II = 1}
\If {$x[i][j] > maxValue$}
    \State $maxValue \gets x[i][j]$
\EndIf
\EndFor
\EndFor
\end{algorithmic}
\end{minipage}
\hspace{-1.4cm}
\begin{minipage}[b]{0.35\linewidth}
\begin{algorithmic}
\Statex{\textbf{Exponential:}}
\For{$(i=1;i<=SL;i++)$}
\State \textbf{\#pragma HLS pipeline off}
\For{$(j=1;j<=SL;j++)$}
\State \textbf{\#pragma HLS pipeline II = 1}
\State $x[i][j] \gets exp(x[i][j]-$
\State $               \ \ \ \ \ \ \ \ \ \ \ \ \ \ maxValue)$
\State $sum \gets sum + x[i][j] $
\EndFor
\EndFor
\end{algorithmic}
\end{minipage}
\hspace{-0.2cm}
\begin{minipage}[b]{0.3\linewidth}
\begin{algorithmic}
\Statex{\textbf{Normalization:}}
\For{$(i=1;i<=SL;i++)$}
\State \textbf{\#pragma HLS pipeline off}
\For{$(j=1;j<=SL;j++)$}
\State \textbf{\#pragma HLS pipeline}
\State \textbf{\ \ \ \ \ \ \ \ \ \ \ \ \ \ \ \ \ \ \ \ \ \ \ \ \ \ \ \ \ II = 1}
\State $x[i][j] \gets \frac{x[i][j]}{sum}$
\EndFor
\EndFor
\end{algorithmic}
\end{minipage}
\end{algorithm}

\begin{algorithm}[H]
\centering
\caption{Layer Normalization}\label{LNorm}
\hspace{-1.0cm}\begin{minipage}[b]{0.5\textwidth} %
\begin{algorithmic}[1]
\Statex{\textbf{Mean:}}
\For{$(i=1;i<=SL;i++)$}
\State \textbf{\#pragma HLS pipeline off}
\For{$(j=1;j<=d_{model};j++)$}
\State \textbf{\#pragma HLS pipeline II = 1}
\State $m[i] \gets m[i] + inputs[i][j]$
\EndFor
    \State $m[i] \gets m[i]/ Embedding\_Dimension; $
\EndFor
\end{algorithmic}
\end{minipage}
\hspace{-1.0cm}\begin{minipage}[b]{0.5\linewidth} %
\begin{algorithmic}[1]
\Statex{\textbf{Variance:}}
\For{$(i=1;i<=SL;i++)$}
\State \textbf{\#pragma HLS pipeline off}
\For{$(j=1;j<=d_{model};j++)$}
\State \textbf{\#pragma HLS pipeline II = 1}
\State $v[i] \gets v[i] + (inputs[i][j]-m[i])^2$
\EndFor
    \State $m[i] \gets m[i]/ Embedding\_Dimension; $
\EndFor
\end{algorithmic}
\end{minipage}
\hfill \\ 
\hfill \\ 
\begin{minipage}[b]{.5\textwidth}
\begin{algorithmic}[1]
\Statex{\textbf{Normalization:}}
\For{$(i=1;i<=SL;i++)$}
\State \textbf{\#pragma HLS pipeline off}
\For{$(j=1;j<=d_{model};j++)$}
\State \textbf{\#pragma HLS pipeline II = 1}
\State $norm_{out}[i][j] \gets \frac{(inputs[i][j]-m[i]}{\sqrt{v[i]+\epsilon}}$
\EndFor
\EndFor
\end{algorithmic}
\hfill
\end{minipage}
\hspace{-1.0cm}\begin{minipage}[b]{0.55\textwidth}
\begin{algorithmic}[1]
\Statex{\textbf{Final Output:}}
\For{$(i=1;i<=SL;i++)$}
\State \textbf{\#pragma HLS pipeline off}
\For{$(j=1;j<=d_{model};j++)$}
\State \textbf{\#pragma HLS pipeline II = 1}
\State $outputs[i][j]\gets gamma[j]\times norm_{out}[i][j]+;$
\State $\ \ \ \ \ \ \ \ \ \ \ \ \ \ \ \ \ \ \ \ \ \ beta[j];$
\EndFor
\EndFor
\end{algorithmic}
\end{minipage}
\end{algorithm}

\begin{minipage}{0.45\textwidth}
\begin{algorithm}[H]
\centering
\caption{Q, K, V Calculation}\label{QKV}
\begin{algorithmic}[1]
\For{$(i=1;i<=Sequence\ Length;i=i+1)$}
\State \textbf{\#pragma HLS pipeline off}
\State $S_q \gets 0$
\State $S_k \gets 0$
\State $S_v \gets 0$
\For{$(k=1;k<=\frac{d_{model}}{h};k++)$}
\State \textbf{\#pragma HLS pipeline II = 1}
\For{$(j=1;j<=\frac{d_{model}}{TS_{MHA}};j++)$}
    \State $S_q \gets S_q + X[i][j]\times W_Q[k][j];$
    \State $S_k \gets S_k + X[i][j]\times W_K[k][j];$
    \State $S_v \gets S_v + X[i][j]\times W_V[k][j];$
\EndFor
    \State $Q[i][k] \gets Q[i][k] + S_q;$
    \State $K[i][k] \gets K[i][k] + S_k;$
    \State $V[i][k] \gets V[i][k] + S_v;$
\EndFor
\EndFor
\end{algorithmic}
\end{algorithm}
\end{minipage}
\hfill \ \ \ \
\begin{minipage}{0.5\textwidth}
\begin{algorithm}[H]
\centering
\caption{FFN3 Calculation}\label{FFN3_alg}
\begin{algorithmic}[1]
\For{$(i=1;i<=Sequence\ Length;i=i+1)$}
\State \textbf{\#pragma HLS pipeline off}
\State $m \gets index \times \frac{Embedding\ Dimension}{Tiles\ in\ FFN}$
\For{$(j=1;j<=\frac{d_{model}}{Tiles\ in\ FFN};j++)$}
\State \textbf{\#pragma HLS pipeline II = 1}
\State $sum \gets 0$
\For{$(k=1;k<=\frac{4\times d_{model}}{Tiles\ in\ FFN};k++)$}
    \State $sum \gets sum + inputs[i][k]\times weights[k][j];$
\EndFor
    \State $output[i][m] \gets output[i][j] + sum;$
    \State $m \gets m+1; $
\EndFor
\EndFor
\end{algorithmic}
\end{algorithm}
\end{minipage}

\begin{minipage}{0.45\textwidth}
\begin{algorithm}[H]
\centering
\caption{$Q\times K^T$ Calculation}\label{QK}
\begin{algorithmic}[1]
\For{$(i=1;i<=SL;i=i+1)$}
\State \textbf{\#pragma HLS pipeline off}
\For{$(j=1;j<=SL;j=j+1)$}
\State \textbf{\#pragma HLS pipeline II = 1}
\State $S \gets 0$
\For{$(k=1;k<=\frac{d_{model}}{h};k++)$}
    \State $S \gets S + Q[i][k]\times K[j][k];$
\EndFor
    \State $s[i][j] \gets S/Embedding\_Dimension;$
\EndFor
\EndFor
\end{algorithmic}
\end{algorithm}
\end{minipage}
\hfill \ \
\begin{minipage}{0.4\textwidth}
\begin{algorithm}[H]
\centering
\caption{$S\times V$ Calculation}\label{SV}
\begin{algorithmic}[1]
\For{$(i=1;i<=SL;i=i+1)$}
\State \textbf{\#pragma HLS pipeline off}
\For{$(j=1;j<=\frac{d_{model}}{h};j++)$}
\State \textbf{\#pragma HLS pipeline II = 1}
\State $vv \gets 0$
\For{$(k=1;k<=SL;k=k+1)$}
    \State $vv \gets vv + S[i][k]\times V[k][j];$
\EndFor
    \State $SV[i][j] \gets vv;$
\EndFor
\EndFor
\end{algorithmic}
\end{algorithm}
\end{minipage}
\begin{algorithm}[H]
\centering
\begin{algorithmic}[1]
\caption{Bias add unit 3}\label{ba3_alg}
\For{$(i=1;i<=Sequence\ Length;i++)$}
\State \textbf{\#pragma HLS pipeline off}
\For{$(j=1;j<=Hidden\ Dimension;j++)$}
\State \textbf{\#pragma HLS pipeline II = 1}
    \State $FF_{out}[i][j] \gets FF_{out}[i][j]+bias_{FFN}[j];$
    \State $FF_{out}[i][j] \gets relu(FF_{out}[i][j]);$
\EndFor
\EndFor
\end{algorithmic}
\end{algorithm}

\begin{minipage}{0.5\textwidth}
\begin{algorithm}[H]
\centering
\caption{FFN1 Calculation}\label{FFN1_alg}
\begin{algorithmic}[1]
\For{$(i=1;i<=Sequence\ Length;i=i+1)$}
\State \textbf{\#pragma HLS pipeline off}
\State $m \gets index \times \frac{Embedding\ Dimension}{Tiles\ in\ FFN}$
\For{$(j=1;j<=\frac{d_{model}}{Tiles\ in\ FFN};j++)$}
\State \textbf{\#pragma HLS pipeline II = 1}
\State $sum \gets 0$
\For{$(k=1;K<=\frac{d_{model}}{Tiles\ in\ FFN};k++)$}
    \State $sum \gets sum + inputs[i][k]\times weights[k][j];$
\EndFor
    \State $output[i][m] \gets output[i][j] + sum;$
    \State $m \gets m+1; $
\EndFor
\EndFor
\end{algorithmic}
\end{algorithm}
\end{minipage}
\hfill
\begin{minipage}{0.5\textwidth}
\begin{algorithm}[H]
\centering
\caption{FFN2 Calculation}\label{FFN2_alg}
\begin{algorithmic}[1]
\For{$(i=1;i<=Sequence\ Length;i++)$}
\State \textbf{\#pragma HLS pipeline off}
\State $m \gets index \times \frac{Hidden\ Dimension}{Tiles\ in\ FFN}$
\For{$(j=1;j<=\frac{4\times d_{model}}{Tiles\ in\ FFN};j++$}
\State \textbf{\#pragma HLS pipeline II = 1}
\State $sum \gets 0$
\For{$(k=1;k<=\frac{d_{model}}{Tiles\ in\ FFN};k++)$}
    \State $sum \gets sum + inputs[i][k]\times weights[k][j];$
\EndFor
    \State $output[i][m] \gets output[i][j] + sum;$
    \State $m \gets m+1; $
\EndFor
\EndFor
\end{algorithmic}
\end{algorithm}
\end{minipage}
\begin{minipage}{0.5\textwidth}
\begin{algorithm}[H]
\centering
\caption{Bias add unit 1}\label{ba1_alg}
\begin{algorithmic}[1]
\For{$(i=1;i<=Sequence\ Length;i=i+1)$}
\State \textbf{\#pragma HLS pipeline off}
\For{$(k=1;k<=\frac{d_{model}}{h};k++)$}
\State \textbf{\#pragma HLS pipeline II = 1}
    \State $Q[i][k] \gets Q[i][k] + bias_q[k];$
    \State $K[i][k] \gets K[i][k] + bias_k[k];$
    \State $V[i][k] \gets V[i][k] + bias_v[k];$
\EndFor
\EndFor
\end{algorithmic}
\end{algorithm}
\end{minipage}
\hfill \ \ \ \
\begin{minipage}{0.5\textwidth}
\begin{algorithm}[H]
\centering
\caption{Bias add unit 2}\label{ba2_alg}
\begin{algorithmic}[1]
\For{$(i=1;i<Sequence\ Length;i++)$}
\State \textbf{\#pragma HLS pipeline off}
\For{$(j=1;j<Embedding\ Dimension;j++)$}
\State \textbf{\#pragma HLS pipeline II = 1}
    \State $FF_{out}[i][j] \gets FF_{out}[i][j]+bias_{FFN}[j];$
\EndFor
\EndFor
\end{algorithmic}
\end{algorithm}
\end{minipage}

\begin{algorithm}
\caption{Software Program}\label{sp}
\begin{algorithmic}[1]
\State Assign the accelerator and other devices with IDs and base addresses
\State Initialize and configure the accelerator and other devices
\State Write to the registers of the configurable parameters: \texttt{Sequence}, \texttt{Heads}, \texttt{Layers\_enc}, \texttt{Layers\_dec}, \texttt{Embeddings}, \texttt{Hidden}, \texttt{Out}
\For{\texttt{i} from 0 to \texttt{no.\_of\_inputs}} \Comment{Iterate based on the number of tiles and layers}
    \State Load input axi master interface buffers with data \Comment{Same tasks for all input interfaces}
 \EndFor
\For{\texttt{i} from 0 to \texttt{no.\_of\_weights}} \Comment{Iterate based on the number of tiles and layers}
    \State Load weight axi master interface buffers with data \Comment{Same tasks for all weight interfaces}
\EndFor
\For{\texttt{i} from 0 to \texttt{no.\_of\_biases}} \Comment{Iterate based on the number of tiles and layers}
    \State Load bias axi master interface buffers with data \Comment{Same tasks for all bias interfaces}
\EndFor
\State Write to control register to start the accelerator 
\State Write to control register to start the timer
\State \textbf{Record} Start time
\While {accelerator is not done}
\State Read status register until the accelerator has finished
\EndWhile
\State \textbf{Record} End time
\State \textbf{Compute} $ Execution\_time \gets End\_time - Start\_time ;$
\end{algorithmic}
\end{algorithm}

\printcredits

\bibliographystyle{IEEEtranN}

\bibliography{cas-refs}

\end{document}